\documentclass[12pt]{article}
\usepackage[utf8]{inputenc}
\usepackage[margin=1in]{geometry}
\usepackage{setspace}
\usepackage{amsmath}
\usepackage{amsfonts}
\usepackage{stackrel}
\usepackage[ruled,vlined]{algorithm2e}
\usepackage{caption}
\usepackage{subcaption}
\usepackage{color}
\newtheorem{definition}{Definition}
\newtheorem{theorem}{Theorem}

\newtheorem{lemma}{Lemma}

\usepackage{hyperref}
\usepackage{float}
\newtheorem{example}{Example}
\newtheorem{corollary}{Corollary}
\usepackage{amssymb}
\usepackage{authblk}

\usepackage{mathtools}
\DeclareMathOperator*{\argmax}{arg\,max}
\DeclareMathOperator*{\argmin}{arg\,min}


\newcommand*{\QEDA}{\null\nobreak\hfill\ensuremath{\square}}

\title{Efficient Carpooling and Toll Pricing for Autonomous Transportation}
\author{Saurabh Amin, Patrick Jaillet, Manxi Wu
\thanks{S. Amin is with the Laboratory for Information and Decision Systems, P. Jaillet is with the Department of Electrical Engineering and Computer Science, Laboratory for Information and Decision Systems, and Operations Research Center, M. Wu is with the Institute for Data, Systems, and Society, Massachusetts Institute of Technology (MIT), Cambridge, MA, USA, \{amins, jaillet, manxiwu\}@mit.edu}}
\date{}
\newcommand{\m}{m}
\newcommand{\M}{M}
\newcommand{\e}{e}
\newcommand{\E}{E}
\newcommand{\qe}{q_e}
\renewcommand{\r}{r}
\newcommand{\R}{R}
\renewcommand{\b}{b}
\newcommand{\B}{B}
\newcommand{\deleq}{\stackrel{\Delta}{=}}
\renewcommand{\(}{\left(}
\renewcommand{\)}{\right)}

\newcommand{\disumbtilde}{\gamma^m(|\tildeb|)}
\newcommand{\xbaropt}{\xbar^{*}}
\newcommand{\thetafun}{\theta}
\newcommand{\unitcost}{\delta}

\newcommand{\xrbbar}{\xbar_r(\bbar)}

\newcommand{\xhatopt}{\xhat^{*}}
\newcommand{\vrmbbar}{v_r^m(\bbar)}
\newcommand{\Ball}{\bar{\B}}
\newcommand{\ball}{\bar{\b}}
\newcommand{\W}{\overline{V}}

\newcommand{\disumbbar}{\gamma^m(|\bbar|)}
\newcommand{\unitcostbbar}{\delta|\bbar|}
\newcommand{\Vbar}{V}
\newcommand{\Vrbbar}{\Vbar_r(\bbar)}
\newcommand{\rep}{h_r}
\newcommand{\bbar}{b}
\newcommand{\etamr}{\eta^m_r}
\newcommand{\muopt}{\mu^*}

\renewcommand{\l}{l}
\renewcommand{\L}{L}
\newcommand{\yopt}{y^*}
\newcommand{\Ropt}{\R^{*}}
\newcommand{\vrmb}{v_{\r}^{\m}(\b)}

\newcommand{\Vrb}{V_\r(\b)}

\newcommand{\x}{x}
\newcommand{\xp}{x^1}
\newcommand{\xpp}{x^2}
\newcommand{\trpp}{t_{\rpp}}
\newcommand{\trp}{t_{\rp}}
\newcommand{\rptwo}{r^{1'}}
\newcommand{\foptp}{f^{1*}}
\newcommand{\foptpp}{f^{2*}}
\newcommand{\fhatp}{\hat{f}^1}
\newcommand{\fhatpp}{\hat{f}^{2}}

\newcommand{\xopthatp}{\xhat^{1*}}
\newcommand{\xopthatpp}{\xhat^{2*}}
\newcommand{\G}{G}
\newcommand{\Lr}{L_r}
\newcommand{\xbar}{x}
\newcommand{\qr}{q_r}

\newcommand{\y}{y}

\newcommand{\bhat}{\hat{b}}
\newcommand{\rhat}{\hat{r}}
\newcommand{\xrb}{\x_r(\b)}
\newcommand{\Wrb}{\W_r(\ball)}

\newcommand{\econ}{\mathcal{G}}
\newcommand{\xrbopt}{\x_r^{*}(\b)}
\newcommand{\xoptmm}{\x^{-\m*}}
\newcommand{\pmopt}{p^{\m*}}
\newcommand{\p}{p}

\newcommand{\popt}{p^*}

\newcommand{\tr}{t_r}
\newcommand{\V}{V}
\newcommand{\te}{t_e}
\renewcommand{\pm}{p^m}
\newcommand{\tripm}{\alpha^m}
\newcommand{\vm}{\beta^m}

\newcommand{\toll}{\tau}
\newcommand{\tolle}{\toll_e}
\newcommand{\um}{u^m}
\newcommand{\Bbar}{B}
\newcommand{\Sx}{S(\x)}
\newcommand{\X}{X}
\newcommand{\pmdag}{p^{\m\dagger}}
\newcommand{\yall}{y}
\newcommand{\augV}{W}

\newcommand{\tildeb}{\rep(\ball)}
\newcommand{\xhat}{\hat{x}}
\newcommand{\xhatp}{\xhat^{1}}
\newcommand{\xhatpp}{\xhat^{2}}

\newcommand{\jhat}{\hat{j}}
\newcommand{\bj}{b_j}

\newcommand{\tmin}{t_{min}}

\newcommand{\tildeE}{\tilde{E}}
\newcommand{\rmin}{r_{min}}
\newcommand{\rphat}{\hat{r}^1}
\newcommand{\tildeR}{\tilde{R}}
\newcommand{\rpphat}{\hat{\r}^2}
\newcommand{\bjhat}{b_{\jhat}}

\newcommand{\lambdar}{\lambda_r}

\newcommand{\umaxm}{\u^{\m\dagger}}
\newcommand{\xopt}{\x^{*}}
\newcommand{\tildeq}{\tilde{q}}
\newcommand{\tildef}{\tilde{f}}
\newcommand{\rp}{r^1}
\newcommand{\rpp}{r^2}
\newcommand{\Gp}{G^1}
\newcommand{\Gpp}{G^2}
\newcommand{\xoptp}{\x^{1*}}
\newcommand{\xoptpp}{x^{2*}}
\newcommand{\Xp}{X^1}
\newcommand{\Xpp}{X^2}
\newcommand{\Rpp}{R^2}
\newcommand{\Rp}{R^1}

\renewcommand{\u}{u}

\newcommand{\K}{K}

\newcommand{\kr}{k_r}

\newcommand{\Uopt}{U^*}

\newcommand{\umax}{u^{\dagger}}

\newcommand{\Bhat}{\hat{B}}
\newcommand{\pdag}{p^{\dagger}}
\newcommand{\udag}{u^{\dagger}}
\newcommand{\tolldag}{\toll^{\dagger}}
\newcommand{\bl}{\ball_l}

\newcommand{\udagm}{u^{\m\dagger}}

\newcommand{\Vmax}{V_{max}}
\newcommand{\phifun}{\phi}
\newcommand{\kopt}{k^{*}}

\newcommand{\fhat}{\hat{f}}
\renewcommand{\xopt}{x^{*}}
\newcommand{\uopt}{u^{*}}
\newcommand{\umopt}{u^{m*}}
\newcommand{\tollopt}{\toll^{*}}

\begin{document}

\maketitle

\begin{abstract}
    In this paper, we address the existence and computation of competitive equilibrium in the transportation market for autonomous carpooling first proposed by \cite{ostrovsky2019carpooling}. At equilibrium, the market organizes carpooled trips over a transportation network in a socially optimal manner and sets the corresponding payments for individual riders and toll prices on edges. The market outcome ensures individual rationality, stability of carpooled trips, budget balance, and market clearing properties under heterogeneous rider preferences. We show that the question of market’s existence can be resolved by proving the existence of an integer optimal solution of a linear programming problem. We characterize conditions on the network topology and riders’ disutility for carpooling under which a market equilibrium can be computed in polynomial time. This characterization relies on ideas from the theory of combinatorial auctions and minimum cost network flow problem. Finally, we characterize a market equilibrium that achieves strategyproofness and maximizes welfare of individual riders.
\end{abstract}
\section{Introduction}\label{sec:introduction}

Autonomous transportation has the potential to significantly transform urban mobility when the technology becomes mature enough for real-world deployment. A significant fleet of driverless cars could be utilized to organize carpooled trips at a much cheaper price and in a more flexible manner relative to the current mobility services that rely on human drivers. Naturally, this technology would reshape the riders' incentives to make trips and share cars. Whether autonomous driving technology will relieve or aggravate congestion crucially depends on how riders will be incentivized to participate in efficient carpooled trips that are constrained by socially optimal tolls. Thus, to fully exploit the potential of self-driving cars, we need to address the complementarity between efficient carpooling and optimal tolling for riders with heterogeneous preferences.\footnote{For example, when toll prices are zero on all roads, all riders will choose to take the shortest route in the network, and the traffic load will exceed the capacity. As the toll prices of edges on this route increase, riders will be incentivized to take carpooled trips in order to split the toll prices (or switch to longer routes).} 

In \cite{ostrovsky2019carpooling}, the authors introduced a competitive market model to study riders' incentives to participate in autonomous carpooled trips and share the road capacity in a socially optimal manner. In this model, the transportation authority sets toll prices on edges, and riders organize carpooled trips and make payments to split the toll prices and trip costs. An outcome is defined by organized trips, riders' payments, and edge tolls. A market equilibrium is defined as an outcome that satisfies the following conditions: {\em individual rationality, stability, budget balance, and market clearing}. The authors show that a market equilibrium may not always exist, but when it does the equilibrium carpooled trips are {\em socially optimal}. This result essentially follows from the first welfare theorem for competitive markets. Our goal in this paper is to address the question of existence of market equilibrium and provide tools to compute a desirable equilibrium. 

Building on \cite{ostrovsky2019carpooling}, we consider that each rider's value of carpooled trips is equal to the value of completed trip, minus the travel time cost and a disutility from carpooling that depends on the size of rider group in the carpool. We make three contributions for this setting: (1) We derive sufficient conditions under which market equilibrium exists; (2) We provide a computational approach to efficiently compute equilibrium outcomes; and (3) We characterize an equilibrium in which riders truthfully report their preferences to a neutral platform that facilitates market implementation. 


Market equilibrium is challenging to analyze because trip organization is essentially a coalition formation problem, in which the riders form carpooled groups and split payments in a manner which ensures that toll prices clear the market. Both trip organization and toll pricing are crucially influenced by the network topology since any trip on a certain route consumes a unit capacity of all edges in that route. Consequently, the toll price on an edge can impact the usage of all edges on the route. The classical methods in mechanism design and coalition games cannot be readily applied to address these features. To address this challenge, we develop a new approach that draw ideas from combinatorial auction theory and network flow optimization. 

We now discuss our approach and main results. In Sec.~\ref{sec:general}, we analyze the linear programming relaxation of the optimal trip organization problem and its dual program. We find that the problem of existence of market equilibrium can be equivalently posed as the problem of existence of an integer optimal solution for the relaxed linear program. Consequently, our goal becomes that of finding conditions under which the primal linear program has an optimal integer solution. Moreover, when these conditions hold, by strong duality of linear programming, optimal solutions of the primal and dual programs provide us an equilibrium outcome of the market. 

We show that when the network is series-parallel and riders have homogeneous levels of carpool disutility, the primal program is guaranteed to have an integer optimal solution, and thus market equilibrium exists (Sec.~\ref{sec:tractable_pooling}). The condition that the {\em network is series-parallel} allows us to compute a set of routes with integer capacities such that the optimal trip organization in the sub-network formed by these routes is also optimal for the original network. These routes can be computed by a greedy algorithm that selects routes in the increasing order of travel time and greedily allocates network capacity to them. Intuitively, the algorithm select routes in a manner that minimizes the total travel time and carpool disutility costs for all trips on a series-parallel network. This intuition however does not apply to non-series-parallel networks. In fact, we provide an example to demonstrate that market equilibrium may not exist in a Wheatstone network. 

On the other hand, the condition that {\em riders have homogeneous carpool disutility} allows us to augment the trip value function of each route into a function that is monotonic in rider groups and satisfies the gross substitutes condition (\cite{gul1999walrasian, de2003combinatorial, leme2017gross}). The augmented trip value function and the route set obtained from aforementioned greedy algorithm can be used to define an equivalent ``economy'', in which the riders are viewed as ``indivisible goods'' and each unit capacity on the routes is viewed as an ``agent''. Using this definition, we show that the existence of market equilibrium on the sub-network is mathematically equivalent to the existence of a Walrasian equilibrium in the economy, which is guaranteed by the gross substitutes condition~\cite{kelso1982job}. 

The issue of equilibrium computation in the autonomous carpooling market is addressed in Sec.~\ref{sec:algorithm}. Our approach for computing optimal carpooled trips takes two steps: firstly, we compute the set of optimal routes using the greedy algorithm; and secondly, we compute the optimal trips on these routes by using well-known Kelso-Crawford algorithm that provides optimal good allocation in Walrasian equilibrium of the equivalent economy. Moreover, riders' equilibrium utilities and toll prices can be computed from the dual linear program using a separation-based method that also relies on the gross substitutes condition. 

Finally, we identify a particular market equilibrium under which riders truthfully report their preferences to a platform (in our context, this is a a neutral entity which facilitates the market implementation). We find that in this equilibrium, riders' payments are equal to their externalities on other riders, and hence are equivalent to the payments in the classical Vickery-Clark-Grove mechanism. This equilibrium also has the advantage of achieving the highest rider utilities among all market equilibria, and only collecting the minimum total toll prices.

\subsection*{Related literature}
\noindent\textbf{Autonomous vehicle market design and competition.}  The paper \cite{siddiq2019ride} studied the impact of competition between two ride-hailing platforms on their choices of autonomous vehicle fleet sizes, prices and wages of human drivers. The authors of \cite{lian2020autonomous} studied the prices in ride-hailing markets, where an uncertain aggregate demand is served by a fixed fleet of autonomous vehicles and elastic supply of human drivers. They argue that the only design that unambiguously reduces the service prices corresponds to the setting when the provision of autonomous carpooled trips occurs in a competitive environment. This finding aligns well with our focus on a competitive autonomous carpooling market. We show that by exploiting the complementarity between carpooling and road pricing, we can achieve an equilibrium outcome that is socially optimal (when sufficient conditions for equilibrium existence are satisfied). 

\noindent\textbf{Human-driven ride-hailing platforms.} A rich body of literature exists on matching and pricing schemes in ride-hailing platforms that rely on supply of human-driven cars. These work includes online matching (\cite{ashlagi2019edge, ozkan2020dynamic}), dynamic and spatial pricing (\cite{banerjee2015pricing, cachon2017role, castillo2017surge, bimpikis2019spatial}), and stochastic control and queuing (\cite{gurvich2015dynamic, afeche2018ride, banerjee2018state}). A key challenge in these problems comes from the two-sided nature of matching between riders and human drivers. In contrast, the autonomous carpooling market that we consider focuses on forming carpooling groups among riders with heterogeneous preferences with constraints imposed by car size, route capacity, edge tolls, and network structure.


\section{A Market Model}\label{sec:model}
\subsection{Network, Riders, and Trips}
Consider a traffic network modeled as a directed graph with a single origin-destination pair. The set of edges in the network is $\E$, and the capacity of each edge $\e \in \E$ is a positive integer $\qe \in \mathbb{N}_{+}$. The set of routes is $\R$, where each route $\r \in \R$ is a sequence of edges that form a directed path from the origin to the destination. We denote the travel time of each edge $\e$ as $\te >0$, and the travel time of each route $\r$ as $\tr= \sum_{\e \in \r} \te$.\footnote{Thus, in our setting, each edge has an L-shaped cost function: cost is a constant when the edge load is below the edge capacity, and becomes extremely high once the load exceeds capacity. In the context of traffic congestion: when the traffic load is below the road capacity, all vehicles pass through the segment at the free-flow speed. However, when the traffic load exceeds the capacity, the travel time significantly increases due to congestion. In our market mechanism, the toll prices are set to ensure that the load of each edge does not exceed its capacity. }

A finite set of riders $\m =1, \dots, \M$ want to take autonomous carpool trips to travel from the origin to the destination. A \emph{trip} is defined as a tuple $\(\b, \r\)$, where $\b$ is the group of riders taking route $\r$ during the trip.\footnote{All individuals in the set $\b$ of an autonomous carpool trip are riders. On the other hand, in human-driven carpool trips, we need to designate a driver in the set $\b$, and match riders with drivers.} The maximum number of riders in any group must be below the capacity of individual car, denoted $A$.\footnote{For simplicity, we assume that cars are of homogeneous capacity.} Thus, the set of rider groups is $\Bbar \deleq \left\{2^{M}\left\vert~ |\b|<A \right.\right\}$, and the set of trips is $ \(\bbar, \r\) \in \Bbar \times \R$. If the group $\b$ in a trip $\(\b, \r\)$ is a singleton set $\{\m\}$, then rider $\m$ takes a solo trip on route $\r$. Otherwise riders in $\b$ share a pooled trip. Each trip $\(\b, \r\)$ occupies a unit capacity for all edges in route $\r$.

The value of each trip $\(\bbar, \r\)$ for a rider $\m \in \b$, denoted as $\vrmbbar$, is given by: \begin{align}\label{eq:m_valuation_trip}
\vrmbbar &= \tripm - \vm \cdot \tr - \disumbbar \cdot \tr, \quad \forall \bbar \in \{\B |\bbar \ni \m\}, \quad \forall \m \in \M, \quad \forall \r \in \R.
\end{align}
Thus, riders have heterogeneous trip values: The parameter $\tripm$ is rider $\m$'s value of arriving at the destination, $\vm$ is rider $\m$'s value of time, and $\disumbbar$ is rider $\m$'s disutility of sharing the pooled trip with rider group of size $|\bbar|$ for a unit travel time. That is, rider $\m$'s value of each trip $\(\b, \r\)$ equals to their value of arriving at the destination nets the cost of trip time and the carpool disutility. 

The carpool disutility $\disumbbar$ represents the rider $\m$'s inconvenience of sharing the vehicle with other riders in the carpool group, potentially due to the need to share space with others and time spent on taking detours and walking to pick-up location. This disutility only depends on the group sizes rather than the identity of riders in the group, riders' values are identical for any two trips $\(\bbar, \r\)$ and $\(\bbar', \r\)$ with the same group sizes (i.e. $|\bbar|=|\bbar'|$) and the same route $\r$. We consider that the carpool disutility $\disumbbar \geq 0$ for all $|\b|=1, \dots, A$, and the disutility of solo trip is zero, i.e. $\gamma^m(1)=0$ for all $m\in \M$. Thus, all riders prefer to take solo trips rather than pooling with other riders. Additionally, the marginal disutility $\disumbbar- \gamma^m(|\b|-1)$ is non-decreasing in the group size $|\b|$ for all $|\b|=2, \dots, A$, i.e. the extra carpool disutility of adding one rider to any trip $\(\b, \r\)$ is non-decreasing in the original trip size $|\b|$.

The cost of each trip includes the fuel charge and the cost of car's wear and tear. We simply assume that the cost of each trip $\(\bbar, \r\) \in \Bbar \times \R$ is $c_r(\bbar)= \unitcostbbar  \tr$,
and $\delta \geq 0$ is cost of driving one rider for a unit travel time. 

The social value of each trip $\(\bbar, \r\)$ is the summation of the trip values for riders in $\bbar$ nets the cost of trip:  
\begin{align}\label{eq:value_of_trip}
    \Vrbbar = \sum_{\m \in \bbar} \vrmbbar - c_r(\bbar) = \sum_{\m \in \bbar} \tripm - \sum_{\m \in \bbar} \vm \tr - \sum_{\m \in \bbar} \disumbbar\tr - \unitcostbbar \tr, ~ \forall \bbar \in \B, \forall \r \in \R. 
\end{align}
\subsection{Market Equilibrium}
We now discuss how an efficient autonomous carpooling market can be organized. A transportation authority sets non-negative toll prices $\toll=\(\tolle\)_{\e \in \E} \in \mathbb{R}_{\geq 0}^{|\E|}$ on edges in the network, where $\tolle$ is the toll price of edge $\e$. Riders form carpool trips. The trip vector is a binary vector $\xbar=\(\xrbbar\)_{\r \in \R, \bbar \in \Bbar} \in \{0,1\}^{|\Bbar|\times|\R|}$, where $\xrbbar=1$ if trip $\(\bbar, \r\)$ is organized and $\xrbbar=0$ if otherwise. A trip vector $\x$ must satisfy the following feasibility constraints:
\begin{subequations}
\begin{align}
    \sum_{\r \in \R}\sum_{\bbar \ni \m} \xrbbar &\leq 1, \quad \forall \m \in \M,  \label{subeq:at_most_one}\\
\sum_{\r \ni \e} \sum_{\bbar \in \Bbar} \xrbbar & \leq \qe, \quad \forall \e \in \E, \label{subeq:edge_capacity}\\
    \xrbbar &\in \{0, 1\}, \quad \forall \bbar \in \Bbar, \quad \forall \r \in \R, \label{subeq:int}
\end{align}
\end{subequations}
where \eqref{subeq:at_most_one} ensures that no rider takes more than 1 trip, and \eqref{subeq:edge_capacity} ensures that the total number of trips that use any edge $\e \in \E$ does not exceed the edge capacity.

Additionally, each rider $\m \in \M$ makes a payment $\pm$ for covering the cost of their trip and the toll prices of the taken edges. The payment vector is $\p = \(\pm\)_{\m \in \M}$. 

An outcome of the carpooling market is represented by the tuple $\(\xbar, \p, \toll\)$. Given any $\(\xbar, \p, \toll\)$, the utility of each rider $\m \in \M$ equals to the value of the trip that $\m$ takes minus the payment:
\begin{align}\label{eq:u_p}
    \um = \sum_{\r \in \R} \sum_{\bbar \ni \m} \vrmbbar \xrbbar - \pm, \quad \forall \m \in \M. 
\end{align}

We next define four properties of the market outcomes, namely \emph{individual rationality}, \emph{stability}, \emph{budget balance}, and \emph{market clearing}. Firstly, an outcome $\(\xbar, \p, \toll\)$ is \emph{individually rational} if riders' utilities are non-negative: 
\begin{align}\label{eq:ir}
    \um \geq 0, \quad \forall \m \in \M.
\end{align}
That is, no rider has an incentive to opt-out of the market.

Secondly, an outcome $\(\xbar, \p, \toll\)$ is \emph{stable} if there is no rider group in $\B$ that can gain higher utilities by organizing trips that are not included in $\x$. Note that the total utility of all riders in any group $\bbar$ for organizing a trip $\(\bbar, \r\)$ cannot exceed the value of the trip minus the toll price for route $\r$, i.e. $\Vrbbar- \sum_{\e \in \r} \tolle$. Thus, a stable market outcome $\(\x, \p, \toll\)$ requires that the total utilities of riders in $\bbar$ obtained using \eqref{eq:u_p} is higher or equal to the total utility that can be obtained from \emph{any} feasible trip $\(\bbar, \r\)$:\footnote{A stable market outcome $\(\xbar, \p, \toll\)$ is Pareto optimal in that no rider's utility can be improved by organizing different trips that are not in $\xbar$ without decreasing the utilities of other riders.}
\begin{align}\label{eq:stability}
    \sum_{\m \in \bbar} \um \geq \Vrbbar- \sum_{\e \in \r} \tolle, \quad \forall \bbar \in \Bbar, \quad \forall \r \in \R. 
\end{align}

 Thirdly, an outcome $\(\xbar, \p, \toll\)$ is \emph{budget balanced} if the total payments of each organized trip is equal to the sum of the toll prices and the cost of the trip; and moreover a rider's payment is zero if they are not part of any organized trip, i.e. 
 \begin{subequations}\label{eq:trip_bb}
\begin{align}
    &\xrbbar=1, \quad \Rightarrow \quad \sum_{\m \in \bbar} \pm = \sum_{\e \in \r} \toll_e + c_r(\bbar), \quad \forall \bbar \in \Bbar, \quad \forall \r \in \R, \label{subeq:bb}\\
    &\xrbbar=0, \quad \forall \r \in \R, \quad \forall \bbar \ni \m, \quad \Rightarrow \quad \pm = 0, \quad \forall \m \in \M. \label{subeq:p_not_assigned}
\end{align}
\end{subequations}

Fourthly, an outcome $\(\xbar, \p, \toll\)$ is \emph{market-clearing} if there are zero tolls on all edges whose capacity limits are not met:
\begin{align}\label{eq:mc}
    \sum_{\r \ni \e} \sum_{\bbar \in \Bbar} \xrbbar  < \qe, \quad \Rightarrow \quad \tolle=0, \quad \forall \e \in \E.
\end{align}

We define market equilibrium as an outcome that satisfies all four properties: 
\begin{definition}\label{def:market_equilibrium}
A market outcome $\(\xopt, \popt, \tollopt\)$ is an equilibrium if it is individually rational, stable, budget balanced and market clearing. 
\end{definition}

The autonomous carpooling market assumes a competitive environment in that riders are free to join any trip and occupies a unit capacity on any route as long as their total payments cover the trip cost and toll prices. From an implementation viewpoint, the process of trip organization and payment can be facilitated by introducing a market platform.\footnote{For simplicity, we assume that this platform is a simple non-strategic market mediator and does not charge a fee for organizing trips. However, a non-negative constant fee can be added to the model without changing the results.} In such an implementation, each rider $\m \in \M$ reports their preference parameters $\(\tripm, \vm, \(\gamma^m(d)\)_{d=1}^{A}\)$ to the platform, and the platform assigns riders to trips according to the trip vector $\xopt$. Then, riders make payments according to $\popt$ to the platform, and the platform pays for the toll prices $\tollopt$ and trip costs on the riders' behalf. When the vector $\(\xopt, \popt, \tollopt\)$ is a market equilibrium, riders follow the trip assigned by the platform, the payments cover the toll prices and trip costs, and toll prices are non-zero only on edges where the load meets the capacity.  \footnote{The computed market equilibrium depends on the reported preference parameters $\(\alpha, \beta, \gamma\)$. For simplicity, we drop the dependence of $\(\xopt, \popt, \tollopt\)$ with respect to these parameters in notation.}

In paper \cite{ostrovsky2019carpooling}, the authors argued that such a transportation market can be mapped into a standard competitive market, where the market equilibrium defined in Definition \ref{def:market_equilibrium} is equivalent to the standard concept of competitive equilibrium. The key issue that we seek to investigate is that market equilibrium may not exist since the edge capacities and riders are indivisible. On the other hand, if an equilibrium $\(\xopt, \popt, \tollopt\)$ exists, from the first welfare theorem, we conclude that the trip vector $\xopt$ necessarily maximizes the total social welfare (Theorem 1 in \cite{ostrovsky2019carpooling}); i.e., $\xopt$ is an optimal solution of the following optimal trip organization problem: 
\begin{equation}\tag{$\mathrm{IP}$}\label{eq:IP}
\begin{split}
    \max_{\xbar} \quad &S(\xbar) = \sum_{\bbar \in \Bbar}\sum_{\r \in \R} \Vrbbar \xrbbar \notag\\
    s.t. \quad & \text{$\xbar$ satisfies \eqref{subeq:at_most_one} -- \eqref{subeq:int},}
    \end{split}
\end{equation}
where $S(\xbar)$ is the social welfare of all trips given by $\xbar$.


\section{Primal and Dual Formulations}\label{sec:general}

In this section, we show that there exists a market equilibrium if and only if the linear relaxation of the optimal trip organization problem \eqref{eq:IP} has integer optimal solutions. We also show that the equilibrium outcomes can be derived from the optimal solutions from the linear relaxation and its dual program.\footnote{All results in this section hold for arbitrary trip values $\V=\(\Vrbbar\)_{\bbar \in \Bbar, \r \in \R}$.}

We first introduce the linear relaxation of \eqref{eq:IP} and its dual formulation. The primal linear program is as follows:  
\begin{subequations}
        \makeatletter
        \def\@currentlabel{$\mathrm{LP}$}
        \makeatother
        \label{eq:LP1bar}
        \renewcommand{\theequation}{$\mathrm{LP}$.\alph{equation}}
    \begin{align}
         \max_{\xbar} \quad &\Sx = \sum_{\bbar \in \Bbar}\sum_{\r \in \R} \Vrbbar \xrbbar, \notag\\
   s.t. \quad  &  \sum_{\r \in \R}\sum_{\b\ni \m} \xrbbar \leq 1, \quad \forall \m \in \M, \label{subeq:LP11}\\
 \quad &\sum_{\r \ni \e} \sum_{\bbar \in \Bbar} \xrbbar  \leq \qe, \quad \forall \e \in \E, \label{subeq:LP12}\\
    &\xrbbar \geq 0, \quad \forall \bbar \in \Bbar, \quad \forall \r \in \R.\label{subeq:LP13}
    \end{align}
\end{subequations}Note that the constraint $\xrbbar \leq 1$ is implicitly included in \eqref{subeq:LP11}, so it is omitted.

By introducing dual variables $\u=\(\um\)_{\m \in \M}$ for constraints \eqref{subeq:LP11} and $\toll = \(\tolle\)_{\e \in \E}$ for constraints \eqref{subeq:LP12}, the dual program of \eqref{eq:LP1bar} can be written as follows: 
\begin{subequations}
        \makeatletter
        \def\@currentlabel{$\mathrm{D}$}
        \makeatother
        \label{eq:D1bar}
        \renewcommand{\theequation}{$\mathrm{D}$.\alph{equation}}
\begin{align}
    \min_{\u, \toll} \quad &U(\u, \toll)= \sum_{\m \in \M} \um+ \sum_{\e \in \E} \qe \tolle \notag \\
s.t. \quad & \sum_{\m \in \bbar} \um +\sum_{\e \in \r} \tolle \geq \Vrbbar, \quad \forall \bbar \in \Bbar, \quad \forall \r \in \R, \label{subeq:D11}\\
        & \um \geq 0, \quad \tolle \geq 0, \quad \forall \m \in \M, \quad \forall \e \in \E. \label{subeq:D12}
\end{align}
\end{subequations}


\begin{theorem}\label{theorem:primal_dual}
A market equilibrium $\(\xbaropt, \popt, \tollopt\)$ exists if and only if \eqref{eq:LP1bar} has an optimal integer solution. Any optimal integer solution $\xbaropt$ of \eqref{eq:LP1bar} is an equilibrium trip  vector, and any optimal solution $\(\uopt, \tollopt\)$ of \eqref{eq:D1bar} is an equilibrium utility vector and an equilibrium toll vector. The equilibrium price vector $\popt$ is given by:  
\begin{align}\label{eq:p}
    \pmopt=\sum_{\r \in \R}  \sum_{\bbar \ni \m} \xrbopt \vrmbbar - \um, \quad \forall \m \in \M.
\end{align}

\end{theorem}


Thus, the question of existence of market equilibrium is equivalent to resolving whether there exists an integer optimal solution for the LP relaxation of the optimal trip  problem. This result follows from the fact that the four properties of market equilibrium, namely individual rationality, stability, budget balance, and market clearing are equivalent to the constraints of \eqref{eq:LP1bar} and \eqref{eq:D1bar}, and the complementary slackness conditions. From strong duality, a market equilibrium exists if and only if the optimality gap between the linear relaxation \eqref{eq:LP1bar} and the integer problem \eqref{eq:IP} is zero. Hence, the linear relaxation \eqref{eq:LP1bar} must have an integer optimal solution, which is the equilibrium trip vector $\xbaropt$.

Theorem \ref{theorem:primal_dual} turns the problem of finding sufficient conditions on the existence of market equilibrium to finding conditions under which \eqref{eq:LP1bar} has optimal integer solutions. Moreover, it enables us to compute market equilibrium as optimal solutions of \eqref{eq:LP1bar} and \eqref{eq:D1bar}. 

As a consequence, we obtain that the total toll prices of shorter routes (routes with lower travel time) must be no less than that of the longer ones (routes with higher travel time). 
\begin{corollary}\label{cor:shorter}
In any market equilibrium $\(\xopt, \popt, \tollopt\)$, for any $\r, \r' \in \R$ such that $\tr \geq t_{r'}$, $\sum_{e \in \r} \tollopt_e \leq \sum_{\e \in \r'} \tollopt_{e}$.
\end{corollary}
This result is intuitive since for all rider groups, taking a shorter route results in a higher trip value than taking a longer route. Therefore, the toll price (which is charged per unit capacity) of shorter routes must be no less than that of longer routes.

\section{Existence of Market Equilibrium}\label{sec:tractable_pooling}
\begin{figure}[ht]
\centering
        \includegraphics[width=0.35\textwidth]{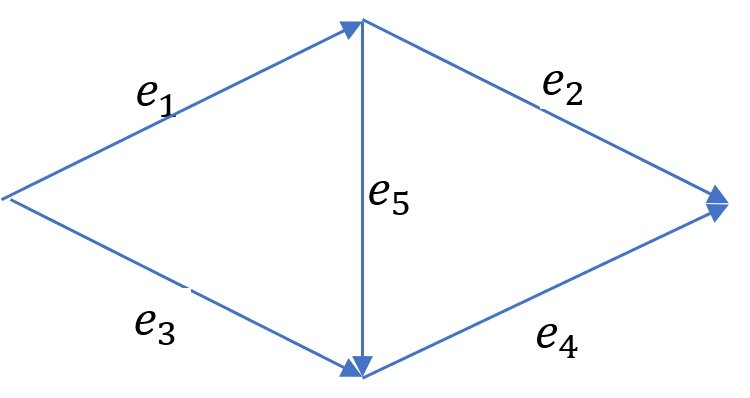}
    \caption{Wheatstone network}
\label{fig:network}
\end{figure}

We characterize the sufficient conditions on network topology and trip values under which the there exists a market equilibrium. We first present an example when market equilibrium does not exist on a wheatstone network. 
\begin{example}\label{ex:wheatstone}
{\normalfont Consider the wheatstone network as in Fig. \ref{fig:network}. The capacity of each edge in the set $\{e_1, e_2, e_3, e_4\}$ is 1, and the capacity of edge $e_5$ is 4. The travel time of each edge is given by $t_1=1$, $t_2=3$, $t_3=3$, $t_4=1$, and $t_5=0$.  

The maximum capacity of vehicle is $A=2$. Three riders $\m=1, 2, 3$ travel on this network. Riders have identical reference parameters: value of trip $\tripm=7$, value of time $\vm=1$, zero carpool disutility, i.e. $\gamma^m(d)=0$ for any $d=1,2$ and any $\m \in \M$, and zero trip cost parameter, i.e.  $\unitcost=0$.

We define the route $e_1$-$e_2$ as $r_1$, $e_1$-$e_5$-$e_4$ as $r_2$, and $e_3$-$e_4$ as $r_3$. Then, trip values are: $V_1(m)=V_3(m)=3$, and $V_2(m)=5$ for all $m \in M$; $V_1(m, m') = V_3(m, m')  =6$, and $V_2(m, m')=10$ for all $m, m' \in \M$.  The unique optimal solution of the linear program \eqref{eq:LP1bar} on this network is $\xopt_1(1, 2)=\xopt_2(2, 3)=\xopt_3(1, 3)=0.5$, and $S(\xopt)=11$. That is, \eqref{eq:LP1bar} does not have an integer optimal solution, and market equilibrium does not exist (Theorem \ref{theorem:primal_dual}). }
\end{example}



We define a network to be series-parallel if a Wheatstone structure as in Example \ref{ex:wheatstone} is not embedded. \begin{definition}[Series-Parallel (SP) Network \cite{milchtaich2006network}]\label{def:series_parallel}
A network is series-parallel if there do not exist two routes that pass through an edge in opposite directions. Equivalently, a network is series-parallel if and only if it is constructed by connecting two series-parallel networks either in series or in parallel for finitely many iterations. 
\end{definition}


Our next theorem shows that market equilibrium is guaranteed to exist if the network is series-parallel (i.e. the Wheatstone structure is not embedded) and riders have homogeneous carpool disutilities.

\begin{theorem}\label{theorem:sp}
Market equilibrium $\(\xopt, \popt, \tollopt\)$ exists if the network is series-parallel and all riders have identical carpool disutility parameters, i.e. 
\begin{align}\label{eq:homogeneous}
    \gamma^m(d)=\gamma(d), \quad \forall d=1, \dots, A, \quad \forall \m \in \M. 
\end{align} 
\end{theorem}

Recall from Theorem \ref{theorem:primal_dual} that showing the existence of market equilibrium is equivalent to proving that \eqref{eq:LP1bar} has an integer optimal solution. Our proof of Theorem \ref{theorem:sp} has three parts: Firstly, we compute an integer route capacity vector $\kopt = \(\kopt_r\)_{\r \in \R}$, where $\Ropt \deleq \{\R|\kopt_r>0\}$ is the set of routes that are assigned positive capacity and $\kopt_r$ is the integer capacity of each route $\r$. We show that when the network is series-parallel, any optimal trip vector for the sub-network with routes $\Ropt$ and capacity vector $\kopt$ is also an optimal trip vector for the original network (Lemma \ref{lemma:FF}). Thus, to prove Theorem \ref{theorem:sp}, we only need to show that there exists an optimal integer solution of trip organization on the sub-network with capacity vector $\kopt$. Secondly, we argue that mathematically the problem of trip organization on the sub-network with capacity vector $\kopt$ can be viewed as a problem of allocating goods in an economy with indivisible goods, and the existence of integer optimal solution is equivalent to the existence of Walrasian equilibrium in the economy (Lemmas \ref{lemma:x_convert_y} -- \ref{lemma:equivalent_economy}). Finally, we show that when riders have homogeneous carpool disutility parameters, the trip value functions satisfy gross substitutes condition. This condition is sufficient to ensure the existence of Walrasian equilibrium in the equivalent economy (Lemmas \ref{lemma:x_convert_y} -- \ref{lemma:integer_kopt}). These three parts ensure that the trip organization problem on the sub-network with capacity vector $\kopt$ has an integer optimal solution, and this solution is also an integer optimal solution of \eqref{eq:LP1bar}. We can thus conclude that a market equilibrium exists.

The rest of this section elaborates on these ideas and presents the lemmas corresponding to each of the three parts. The proofs of these lemmas are included in Appendix \ref{apx:proof_A}.  
\vspace{0.2cm}

\noindent\underline{Part 1}. We first compute the route capacity vector $\kopt$ by a greedy algorithm (Algorithm \ref{alg:flow}). The algorithm begins with finding a shortest route of the network $\rmin$, and sets its capacity as $\kopt_{\rmin} = \min_{\e \in \rmin} \qe$, which is the maximum possible capacity that can be allocated to $\rmin$. After allocating the capacity $\kopt_{\rmin}$ to route $\rmin$, the residual capacity of each edge on $\rmin$ is reduced by $\kopt_{\rmin}$. We then repeat the process of finding the next shortest route and allocating the maximum possible capacity to that route until there exists no route with positive residual capacity in the network. 

Note that in each step of Algorithm \ref{alg:flow}, the capacity of at least one edge is fully allocated to the route that is chosen in that step. Therefore, the algorithm must terminate in less than $|\E|$ number of steps. The algorithm returns the capacity vector $\kopt$, where $\Ropt = \{\R|\kopt_r>0\}$ is the set of routes allocated with positive capacity, and the capacity of each $\r \in \Ropt$ is $\kopt_r$. The remaining routes in $\R\setminus \Ropt$ are set with zero capacity. Since the network is series-parallel, the total capacity given by the output of the greedy algorithm equals to the network capacity $C$ (\cite{bein1985minimum}), i.e. $\sum_{\r \in \Ropt} \kopt_r = C$. 

Moreover, the shortest path of the network in each step can be computed by Dijkstra's algorithm with time complexity of $O(|N|^2)$, where $|N|$ is the number of nodes in the network. Therefore, Algorithm \ref{alg:flow} has time complexity of $O(|N|^2|E|)$.

\begin{algorithm}[ht]
\SetAlgoLined
\textbf{Initialize:} Set $\tildeq_e \leftarrow \qe, ~ \forall \e \in \E$; $\kr \leftarrow0, ~\forall \r \in \R$; $\tildeE \leftarrow \E$; $\(\tmin, \rmin\) \leftarrow ShortestRoute(\tildeE)$\;
\While{$\tmin < \infty$}{
$\kopt_{\rmin} \leftarrow \min_{\e \in \rmin} \tildeq_e$\;
\For{$\e \in \rmin$}{
$\tildeq_e \leftarrow \tildeq_e - \kopt_{\rmin}$\;
\If{$\tildeq_e = 0$}{
$\tildeE \leftarrow \tildeE \setminus \{\e\}$\;
}
}
$\(\tmin, \rmin\) \leftarrow ShortestRoute(\tildeE)$\;
}
\textbf{Return $\kopt$}
\caption{Greedy algorithm for computing route capacity}
\label{alg:flow}
\end{algorithm}


Next, we consider the sub-network comprised of routes in $\Ropt$ with corresponding route capacities given by $\kopt$. Analogous to \eqref{eq:LP1bar}, the linear relaxation of optimal trip organization problem on this sub-network is given by: 
\begin{subequations}
        \makeatletter
        \def\@currentlabel{LP$k^{*}$}
        \makeatother
        \label{eq:LP2k}
        \renewcommand{\theequation}{LP$k^{*}$.\alph{equation}}
        \begin{align}
         \max_{\x} \quad &\Sx = \sum_{\b \in \Bbar}\sum_{\r \in \R} \Vrb \xrb, \notag\\
   s.t. \quad  &  \sum_{\r \in \R}\sum_{\b \ni \m} \xrb \leq 1, \quad \forall \m \in \M, \label{subeq:LP2k1}\\
 \quad & \sum_{\b \in \B} \xrb  \leq \kopt_r, \quad \forall \r \in \R, \label{subeq:LP2k2}\\
    &\xrb \geq 0, \quad \forall \b \in \B, \quad \forall \r \in \R,\label{subeq:LP2k3}
    \end{align}
    \end{subequations}
where \eqref{subeq:LP2k1} ensures that each rider is in at most one trip, and \eqref{subeq:LP2k2} ensures that the total number of trips in each route $\r$ does not exceed the route capacity $\kopt_r$ given by $\kopt$. 
    

\begin{lemma}\label{lemma:FF}
If the network is series-parallel, then any optimal solution of \eqref{eq:LP2k} is an optimal solution of \eqref{eq:LP1bar}.  
\end{lemma}
To prove Lemma \ref{lemma:FF}, we first prove that any feasible solution of \eqref{eq:LP2k} is also a feasible solution of \eqref{eq:LP1bar} by showing that the capacity vector $\kopt$ computed from Algorithm \ref{alg:flow} satisfies $\sum_{\r \ni \e}\kopt_r \leq \qe$ for all $\e \in \E$. Thus, the optimal value of \eqref{eq:LP2k} is no higher than that of \eqref{eq:LP1bar}. 

Next, we argue that for a series-parallel network, the optimal value of \eqref{eq:LP2k} is no less than that of \eqref{eq:LP1bar}; hence, any optimal solution of \eqref{eq:LP2k} must also be an optimal solution of \eqref{eq:LP1bar}. To prove this argument, we show that for any optimal solution $\xhat^*$ of \eqref{eq:LP1bar}, we can construct another trip vector $\xopt$ such that $\xopt$ is feasible in \eqref{eq:LP2k}, and $S(\xopt) \geq S(\xhat^*)$. Such a vector $\xopt$ can be constructed from $\xhat^*$ by re-assigning rider groups $\bhat \in \hat{B} \deleq \{\B|\sum_{\r \in \R}\xhat_r(\bhat)>0\}$ -- the set of rider groups with positive weights in $\xhat$ -- to routes in $\kopt$. Then, for each $\bhat \in \Bhat$, the trip value $\(\bhat, \r\)$ as in \eqref{eq:value_of_trip} can be written as $V_r(\bhat)= \sum_{\m \in \bhat} \tripm - g(\bhat) \tr$, where $g(\bhat) = \sum_{\m \in \bhat} \vm + \sum_{\m \in \bhat} \gamma^m(|\bhat|)+  \unitcost|\bhat|$ is each group $\bhat$'s sensitivity to route travel time. 

Moreover, we define the weight of each group $\bhat$ under the vector $\xhat^*$ as $f(\bhat)=\sum_{\r \in \R} \xhat^*_r(\bhat)$. To construct the new vector $\xopt$, we start with re-assigning weights of the rider groups in $\Bhat$ one-by-one in decreasing order of their sensitivities to the shortest route in $\Ropt$ until the capacity of the shortest route given $\kopt$ is fully utilized. Then, we proceed to assign the weights of the remaining rider groups in $\Bhat$ to the second shortest route in $\Ropt$. This process is repeated until either all weights of rider groups in $\Bhat$ are re-assigned to routes or all routes' capacities in $\kopt$ are used-up. Since the total weight of $\xhat$ is less than or equal to the network capacity $C$, and the total capacity given by $\kopt$ equals to $C$, all weights of $\Bhat$ given by $\xhat$ must get assigned to routes in $\Ropt$ when the algorithm terminates. Additionally, the constructed trip vector $\xopt$ is a feasible solution of \eqref{eq:LP2k}.  

This re-assignment process enables rider groups with higher sensitivity of travel time to take shorter routes. 
This ensures that the constructed $\xopt$ satisfies the inequality $S(\xopt) \geq S(\xhat)$ when the network is series-parallel. We prove this by mathematical induction: First, $S(\xopt) \geq S(\xhat)$ holds trivially on any single link network. Second, if this inequality holds on any two series-parallel networks, then it also holds on the network that is constructed by connecting the two sub-networks in series or in parallel. Since any series-parallel network is constructed by connecting single link networks in series or in parallel for a finite number times, $S(\xopt) \geq S(\xhat)$ must hold for any series-parallel network. Hence, we can conclude that the optimal value of \eqref{eq:LP2k} is no less than that of \eqref{eq:LP1bar}, and any optimal solution of \eqref{eq:LP2k} must also be an optimal solution of \eqref{eq:LP1bar}.

In part 1, Lemma \ref{lemma:FF} ensures that if \eqref{eq:LP2k} has an integer optimal solution, then that solution must be an optimal integer solution of \eqref{eq:LP1bar}. It remains to show that \eqref{eq:LP2k} indeed has an integer optimal solution. 

\vspace{0.2cm}

\noindent\underline{Part 2.} In this part, we first construct an augmented trip value function that is monotonic in the rider group. Then, we construct an auxiliary network comprised of parallel routes with unit capacities based on the set of routes given by $\kopt$. We show that \eqref{eq:LP2k} has an integer optimal solution if and only if the linear relaxation of the trip organization problem on the auxiliary network with the augmented value function has integer optimal solution. Moreover, the trip organization problem on the auxiliary network with the augmented value function is equivalent to an allocation problem in an economy with indivisible goods. The existence of optimal integer solution is equivalent to the existence of Walrasian equilibrium in this economy. 

To begin with, we introduce the definition of monotonic trip value function as follows:
\begin{definition}[Monotonicity]\label{eq:monotonicity}
For each $\r \in \R$, the trip value function $\Vbar_r$ is monotonic if for any $\bbar, \bbar' \in \Bbar$,   $\Vbar_r(\bbar\cup \bbar') \geq \Vbar_r(\bbar)$.
\end{definition}
Monotonicity condition requires that adding any rider group $\bbar'$ to a trip $\(\bbar, \r\)$ does not reduce the trip's value. The monotonicity condition may not be always satisfied in general because of two reasons: First, if the size of riders $|\bbar \cup \bbar'| > A$, then the trip $\(\b\cup\b', \r\)$ is infeasible, and the trip value is not defined. Second, even when $|\bbar \cup \bbar'| \leq A$, the value $V_r(\bbar \cup \bbar')$ may be less than $V_r(\bbar)$ when the carpool disutility is sufficiently high.


We augment $\Vbar: \Bbar \times \R \to \mathbb{N}$ to a monotonic value function $\W: \Ball \times \R \to \mathbb{N}$, where $\Ball \deleq 2^{\M}$ is the set of all rider subsets (including the rider subsets with sizes larger than A). The value of $\W_r(\ball)$ can be written as follows: \begin{align}\label{eq:Vtilde}
    \Wrb \deleq \max_{\bbar \subseteq \ball, ~ \bbar \in \Bbar} \Vbar_r(\bbar),  \quad \forall \r \in \R, \quad \forall \ball \in \Ball. 
\end{align}
That is, the value of any rider group $\ball \in \Ball$ on route $\r$ equals to the maximum value of a feasible trip $\(\bbar, \r\)$ where rider group $\bbar$ is a subset of $\ball$. The augmented value function $\W$ satisfies the monotonicity condition. 

We refer $\tildeb \deleq \argmax_{\bbar \subseteq \ball, ~ \bbar \in \Bbar} V_r(\bbar)$ as the \emph{representative rider group} of $\ball$ for route $\r$. From \eqref{eq:value_of_trip}, we can re-write the augmented trip value function $\W$ as a linear function of travel time: 
\begin{align}\label{eq:value_of_trip_tilde}
    \Wrb =  \sum_{\m \in \tildeb} \tripm - \sum_{\m \in \tildeb} \vm \tr - \sum_{\m \in \tildeb} \disumbtilde \tr - \delta |\tildeb| \tr, ~ \forall \ball \in \Ball, \forall \r \in \R.
\end{align}

Next, we construct an auxiliary network given the set of routes $\Ropt$ with capacity vector $\kopt$ output from Algorithm \ref{alg:flow}. Specifically, we convert each route $\r \in \Ropt$ with integer capacity $\kopt_r$ to the same number of parallel routes each with a unit capacity in the auxiliary network. We denote the route set of the auxiliary network as $\L= \cup_{\r \in \Ropt} \Lr$, where each set $\Lr$ is the set of routes converted from route $\r$ in the original network.

We now consider the trip organization problem on the auxiliary network with the augmented trip value function. For each $\l \in \L$ and each $\ball \in \Ball$, we define $\(\ball, \l\)$ as an augmented trip. In this trip, the rider group $\tildeb$ takes route $\l$ of the auxiliary network, while the remaining riders $\m \in \ball \setminus \tildeb$ are not included in the trip. We denote the augmented trip vector as $\yall = \(\yall_l(\ball)\)_{\ball \in \Ball, \l \in \K} \in \{0, 1\}^{|\Ball| \times \L}$, where $\yall_l(\ball)=1$ if the augmented trip  $\(\ball, \l\)$ is organized, and $\yall_l(\ball)=0$ if otherwise. The value of the augmented trip is defined as $\augV_l(\ball) = \W_r(\ball)$ for any $\ball \in \Ball$, any $l \in \Lr$ and any $\r \in \Ropt$.

For any $\yall \in \{0, 1\}^{|\Ball| \times \L}$, we can compute a trip vector for the original optimal trip organization problem $\x= \chi(\yall) \in \{0,1\}^{|\B| \times \R}$ such that the actually organized trips given by $\x = \chi(\yall)$ are the same as that given by $\yall$. In particular, for each route $\r \in \Ropt$, and each augmented trip $\(\ball, \l\) \in \Ball \times \Lr$ such that $\yall_l(\ball)=1$, we choose a representative rider group $\bhat \in \tildeb$ and set $x_r(\bhat)=1$ for the original trip $\(\bhat, \r\)$ that represents the organized augmented trip $\(\ball, \l\)$.  We set $\xrb=0$ for all other trips. The trip vector $\x=\chi(\yall)$ can be written as follows: 
\begin{align}\label{eq:chi}
    \forall \r \in \Ropt, ~ \forall\(\ball, \l\)  s.t. \yall_l(\ball)=1, ~ \exists \bhat \in \tildeb, ~ s.t. ~ x_r(\bhat)=1, ~\text{and} ~ x_r(\b)=0, ~ \forall \b \in \B \setminus \{\bhat\}
\end{align}

Hence, we write the linear relaxation of optimal trip organization problem on the auxiliary network with the augmented trip value function as follows: 
\begin{subequations}
        \makeatletter
        \def\@currentlabel{LP-y}
        \makeatother
        \label{eq:opt_lane}
        \renewcommand{\theequation}{LP-y.\alph{equation}}
\begin{align}
         \max_{\yall} \quad &S(y) = \sum_{\ball\in \Ball}\sum_{\l \in \L} \augV_l(\ball) y_l(\ball),\notag\\
   s.t. \quad  &  \sum_{l \in L}\sum_{\ball \ni \m} y_l(\ball) \leq 1, \quad \forall \m \in \M,\label{eq:lane_1}\\
 \quad & \sum_{\ball \in \Ball} y_l(\ball)  \leq 1,  \quad \forall l \in L, \label{eq:lane_2}\\
    &y_l(\ball) \geq 0, \quad \forall \ball \in \Ball, \quad \forall l \in L,
\end{align}
\end{subequations}

\begin{lemma}\label{lemma:x_convert_y}
The linear program \eqref{eq:LP2k} has an integer optimal solution if and only if \eqref{eq:opt_lane} has an integer optimal solution. Moreover, if $\yopt$ is an integer optimal solution of \eqref{eq:opt_lane}, then $\xopt = \chi(\yopt)$ as in \eqref{eq:chi} is an optimal integer solution of \eqref{eq:LP2k}. \end{lemma}

This lemma shows that finding an optimal integer solution of \eqref{eq:LP2k} is equivalent to finding an optimal integer solution of \eqref{eq:opt_lane}.

We finally show that the augmented trip organization problem is mathematically equivalent to an economy $\econ$ with indivisible goods, and the existence of market equilibrium in our carpooling market is equivalent to the existence of Walrasian equilibrium of the economy. In $\econ$, the set of indivisible ``goods" is the rider set $\M$ and the set of agents is the route set $\L$ in the auxiliary network. Each agent $\l$'s value of any good bundle $\ball \in \Ball$ is equivalent to the augmented trip value function $W_l(\ball)$. Moreover, each good $\m$'s price is equivalent to rider $\m$'s utility $\um$. The vector of good allocation is $\yall$, where $\yall_l(\ball)=1$ if good bundle $\ball$ is allocated to agent $\l$. Given any $\yall$, for each $\l \in \L$, we denote the bundle of goods that is allocated to $\l$ as $\bl$, i.e. $\yall_l(\bl)=1$. If no good is allocated to $\l$ (i.e. $\sum_{\ball \in \Ball} \yall_l(\ball)=0$), then $\bl=\emptyset$. The Walrasian equilibrium of economy $\econ$ is defined as follows:

\begin{definition}[Walrasian equilibrium \cite{kelso1982job}]
A tuple $\(\yall^*, \uopt\)$ is a Walrasian equilibrium if 
\begin{itemize}
    \item[(i)] For any $\l \in \L$, $\bl \in \argmax_{\ball \in \Ball} W_l(\ball) - \sum_{\m \in \bl} \um$, where $\bl$ is the good bundle that is allocated to $\l$ given $\yopt$
    \item[(ii)] For any $\m \in \M$ that is not allocated to any agent, (i.e. $\sum_{\l \in \L} \sum_{\ball \ni \m} \yopt_l(\ball) =0$), $\umopt=0$. 
\end{itemize}
\end{definition}

In fact, we can show that \eqref{eq:opt_lane} has integer optimal solution if and only if Walrasian equilibrium exists in this equivalent economy: 
\begin{lemma}\label{lemma:equivalent_economy}
The linear program  \eqref{eq:opt_lane} has integer optimal solution if and only if a Walrasian equilibrium $\(\yopt, \uopt\)$ exists in the equivalent economy. Furthermore, $\yopt$ is an integer optimal solution of \eqref{eq:opt_lane}, and $\xopt= \chi(\yopt)$ as in \eqref{eq:chi} is an optimal integer solution of \eqref{eq:LP2k}. 
\end{lemma}

In part 2, from Lemmas \ref{lemma:x_convert_y} -- \ref{lemma:equivalent_economy}, we turn the problem of proving the existence of integer optimal solution in \eqref{eq:LP2k} to proving that the equivalent economy $\econ$ has Walrasian equilibrium.

\vspace{0.2cm}
\noindent\underline{Part 3.} In this final part, we show that if the carpool disutility parameter $\gamma^m$ is homogeneous across all $\m \in \M$, then Walrasian equilibrium exists in the economy $\econ$ constructed in Part 2. 

To begin with, we introduce the following definition of gross substitutes condition on the augmented value function $\W$. In this definition, we utilize the notion of marginal value function $\W_r(\ball'|\ball) = \W_r(\ball \cup \ball') - \W_r(\ball)$ for all $\r \in \R$ and all $\ball, \ball' \subseteq \M$.  

\begin{definition}[Gross Substitutes \cite{reijnierse2002verifying}]\label{def:gross_substitute}
For each $\r \in \R$, the augmented trip value function $\W_r$ is said to satisfy gross substitutes condition if
\begin{itemize}
    \item[(a)] For any $\ball, \ball'\subseteq \Ball$ such that $\ball \subseteq \ball'$ and any $i \in \M \setminus \ball'$, $\W_r(i|\ball') \leq \W_r(i|\ball)$.
    \item[(b)] For all groups $\ball \in \Ball$ and any $i,j,k \in \M \setminus \ball$, 
    \begin{align}\label{eq:gs}
        \W_r(i, j|\ball) + \W_r(k|\ball) \leq \max \left\{\W_r(i|\ball) + \W_r(j,k|\ball), ~ \W_r(j|\ball) + \W_r(i,k|\ball)\right\}.
    \end{align}
\end{itemize}
\end{definition}

In Definition \ref{def:gross_substitute}, \emph{(a)} requires that the augmented value function $\W$ is submodular, i.e. the marginal valuation of $\(\ball, \r\)$ decreases in the size of group $\ball$. Additionally, the gross substitutes condition also requires that the augmented value function satisfy \emph{(b)}. This condition ensures that the sum of marginal values of $\{i, j\}$ and $k$ is not strictly higher than that of both $i, \{j, k\}$ and $j, \{i,k\}$. 

The following lemma shows that when all riders have a homogeneous carpool disutility, the augmented trip value function $\W$ satisfies gross substitutes condition. 
\begin{lemma}\label{lemma:condition_gross}
The augmented value function $\W_r$ satisfies gross substitutes for all $\r \in \R$ if riders have homogeneous carpool disutility: $\gamma^m(d)\equiv \gamma(d)$ for all $d=1, \dots, A$ and all $\m \in \M$.
\end{lemma}

In the economy $\econ$, since each agent $\l$'s value function $W_l(\ball) = \W_r(\ball)$ for all $\ball \in \Ball$ and all $\l \in \Lr$, the agents' value functions $W$ satisfy gross substitutes under the condition in Lemma \ref{lemma:condition_gross}. Moreover, from \eqref{eq:value_of_trip_tilde}, the value functions $W$ are also monotonic. From the following result, we know that a Walrasian equilibrium exists in economy with value functions that satisfy monotonicity and gross substitutes conditions.

\begin{lemma}[\cite{bikhchandani1997competitive}]\label{lemma:parallel}
If $W_l$ satisfies the monotonicity and gross substitutes conditions for all $l \in L$, then Walrasian equilibrium $\(\yopt, \uopt\)$ exists. 
\end{lemma}

Based on Lemmas \ref{lemma:equivalent_economy}, \ref{lemma:condition_gross} and \ref{lemma:parallel}, we conclude the following:  
\begin{lemma}\label{lemma:integer_kopt}
The linear program \eqref{eq:LP2k} has an optimal integer solution if all riders have homogeneous carpool disutilities, i.e. $\gamma^m(d) \equiv \gamma(d)$ for all $\m \in \M$ and all $d=1, \dots, A$.
\end{lemma}

Lemma \ref{lemma:integer_kopt} shows that \eqref{eq:LP2k} has an optimal integer solution. From Lemma \ref{lemma:FF}, we know that this solution is also an optimal integer solution of \eqref{eq:LP1bar}. Therefore, we can conclude Theorem \ref{theorem:sp} that market equilibrium exists when the network is series parallel and riders have homogeneous carpool disutilities. In Sec. \ref{sec:algorithm} and \ref{sec:strategyproof}, we assume that the sufficient conditions in Theorem \ref{theorem:sp} hold, and market equilibrium exists. 
\section{Computing Market Equilibrium}\label{sec:algorithm}

In this section, we present an algorithm for computing the market equilibrium $\(\xopt, \popt, \tollopt\)$. The ideas behind the algorithm are based on Theorems \ref{theorem:primal_dual} -- \ref{theorem:sp} and their proofs.

\vspace{0.2cm}

\noindent\textbf{Computing optimal trip vector $\xopt$.} To begin with, one can obtain the optimal trip vector $\xopt$ following the proof of Theorem \ref{theorem:sp}. In particular, we compute the route capacity vector $\kopt$ from Algorithm \ref{alg:flow}. From Lemma \ref{lemma:FF}, we know that the optimal trip assignment vector $\xopt$ is an optimal integer solution of \eqref{eq:LP2k}. Moreover, from Lemmas \ref{lemma:x_convert_y} -- \ref{lemma:integer_kopt}, we know that: (i) $\xopt$ can be derived from optimal solution $\yopt$ on the auxiliary network with the augmented trip value function $W$; and (ii) $\yopt$ is the same as the optimal good allocation in Walrasian equilibrium of the equivalent economy $\econ$. We introduce the following well-known Kelso-Crawford algorithm (Algorithm \ref{alg:KC}) for computing Walrasian equilibrium $\yopt$.

 \begin{algorithm}[htp]
\textbf{Initialize:} Set $u^m \leftarrow 0 ~ \forall m \in M$; $\bl \leftarrow \emptyset, ~ \forall l \in L$\;
\While{TRUE}{
\For{$\l \in \L$}{$J_l \leftarrow \arg\max_{J \subseteq \M\setminus \bl} \phifun_l(J|\bl) \deleq \left\{W_l(J \cup \bl) - \sum_{m \in \bl} u^m-  \sum_{m \in J} \(u^m+\epsilon\)\right\}$}
\eIf{$J_l =\emptyset, ~ \forall l \in L$}{
break}{Arbitrarily pick $\hat{l}$ with $J_{\hat{l}} \neq \emptyset$\;
$\bar{b}_{\hat{l}} \leftarrow \bar{b}_{\hat{l}} \cup J_{\hat{l}}$\;
$\bar{b}_{\hat{l}} \leftarrow \bar{b}_{\hat{l}} \setminus J_{\hat{l}}, ~ \forall l \neq \hat{l}$\;
$u^m \leftarrow u^m+\epsilon, ~ \forall \m \in J_{\hat{l}}$.}
}
\textbf{Return $\(\bl\)_{l \in L}$}
\caption{Kelso-Crawford Auction \cite{kelso1982job}}
\label{alg:KC}
\end{algorithm}
Algorithm \ref{alg:KC} begins with all riders having zero utilities $\um=0$ and all routes in the auxiliary network being empty $\bl = \emptyset$. In each iteration, we compute the set of riders $J_l$ who are currently unassigned to route $\l$ and maximize the function $\phifun_l(J_l|\bl)$. The function $\phifun_l(J_l|\bl)$ equals to the trip value minus the riders' utilities when the set $J_l$ is added to $\bl$. If there exists a route $\hat{l} \in \L$ with $J_{\hat{l}} \neq \emptyset$, then we assign riders in $J_{\hat{l}}$ to one of such route $\hat{l}$, and increase the utilities of these riders by a small number $\epsilon$.

Algorithm \ref{alg:KC} terminates when $J_l = \emptyset$ for all $\l \in \L$. Given any $\epsilon< \frac{1}{2|\M|}$, when the algorithm terminates, all routes are assigned with the rider set that maximizes its trip value minus riders' utilities. The trip vector based on $\(\bl\)_{l \in L}$ is given by:  
\begin{align}\label{eq:yopt}
    \yopt_l(\bl)= 1, \quad \text{ and } \quad \yopt_l(\ball)=0, \quad \forall \ball \in \Ball \setminus \{\bl\}, \quad \forall \l \in \L. 
\end{align}
The following lemma shows that $\yopt$ is optimal under the conditions of monotonicity and gross substitutes. 
\begin{lemma}[\cite{kelso1982job}]\label{lemma:terminate}
For any $\epsilon< \frac{1}{2|\M|}$, if the augmented value function $W$ satisfies monotonicity and gross substitutes condition, then $\yopt$ as in \eqref{eq:yopt} is an optimal integer solution of \eqref{eq:opt_lane}.
\end{lemma}

Recall from Lemma \ref{lemma:condition_gross}, we know that when all riders have identical carpool disutility, i.e. $\gamma^m(d)=\gamma(d)$ for all $d=1, \dots, A$, then the augmented trip value function $\W$ satisfies gross substitutes condition. Since $W_l(\ball)=\W_r(\ball)$ for all $\l \in \Lr$ and all $\r \in \R$, $W$ also satisfies gross substitutes condition. Therefore, $\yopt$ is a Walrasian equilibrium good allocation vector in the equivalent economy $\econ$, and from Lemmas \ref{lemma:FF} -- \ref{lemma:equivalent_economy}, the vector $\xopt=\chi(\yopt)$ as in \eqref{eq:chi} is an optimal trip vector in market equilibrium.

In each iteration of Algorithm \ref{alg:KC}, we need to compute the set $J_l \in \arg\max_{J \subseteq \M\setminus \bl} \phifun_l(J_l|\bl)$ for each $\l \in \L$. Since the value function $W_l(\ball)$ is monotonic and satisfies gross substitutes condition, $J_l$ can be computed by a greedy algorithm, in which riders are added to the set $J_l$ one by one in decreasing order of the difference between the rider's marginal trip value $W_l(\m|\bl \cup J_l) = W_l(\{\m\} \cup \bl \cup J_l)- W_l(\bl \cup J_l)$ and their utility $\um$ (\cite{kelso1982job}). Since $W_l(\ball)=\W_r(\ball)$ as in \eqref{eq:Vtilde}, and all riders have identical carpool disutility parameter, we can write $W_l(\ball)$ as follows: \begin{align*}
    W_l(\ball)=\Wrb = \sum_{\m \in \tildeb} \etamr -\thetafun(|\tildeb|) \tr, \quad \forall \l \in \Lr, \quad \forall \ball \in \Ball,
\end{align*}
where $\etamr \deleq \tripm - \vm\tr$ and $\thetafun(|\tildeb|)= |\tildeb|\gamma(|\tildeb|) + \unitcost|\tildeb|$. The representative rider group $\rep(\ball)$ for any trip $\(\ball, \r\) \in \Ball \times \R$ can be constructed by selecting riders from $\ball$ in decreasing order of $\etamr$. The last selected rider $\hat{\m}$ (i.e. the rider in $\rep(\ball)$ with the minimum value of $\etamr$) satisfies: 
\begin{align*}
    \eta_r^{\hat{\m}} \geq \(\thetafun(|\rep(\ball)|) - \thetafun(|\rep(\ball)|-1)\)\tr.
\end{align*}
That is, adding rider $\hat{\m}$ to the set $\rep(\ball) \setminus \{\hat{\m}\}$ increases the trip value. Additionally, 
\begin{align*}
    \etamr < \(\thetafun(|\rep(\ball)|+1) - \thetafun(|\rep(\ball)|)\)\tr, \quad \forall \m \in \ball \setminus \rep(\ball).
\end{align*}
We can compute the set $J_l \leftarrow \arg\max_{J \subseteq \M\setminus \bl} \phifun_l(J|\bl)$ in each iteration of Algorithm \ref{alg:KC} using Algorithm \ref{alg:demand}. In this algorithm, we first compute the size of the representative rider group $\tilde{h}= |\rep(\bl)|$, then we add riders not in $\bl$ into $J_l$ greedily according to their marginal trip value minus utility. Note that for computing marginal trip value, we do not need to compute the augmented trip value function $W_l(\ball)$, but simply need to keep track of the representative rider group size $\tilde{h}$. 

 \begin{algorithm}[H]
\textbf{Initialize:} Set $J_l \leftarrow \emptyset$, $\tilde{h}\leftarrow 0$, $\tilde{b}_l \leftarrow \bl$\;
\While{TRUE}{
$\hat{m} \leftarrow \argmax_{\m \in\tilde{b}_l} \eta_l^m$\;
\eIf{$\eta_l^{\hat{m}} < \(\thetafun(\tilde{h}+1) - \thetafun(\tilde{h})\)t_l$}{break}{$\tilde{h} \leftarrow \tilde{h}+1$, $\tilde{b}_l \leftarrow \tilde{b}_l \setminus \{\hat{\m}\}$}
}
\While{TRUE}{
$\hat{j} \leftarrow \argmax_{j \in S\setminus \(\bl \cup J_l\)} \eta_l^j-u^j$\;
\eIf{$\eta_l^j-u^j < \(\thetafun(\tilde{h}+1) - \thetafun(\tilde{h})\)t_l$}{break}{$\tilde{h} \leftarrow \tilde{h}+1$, $J_l \leftarrow J_l \cup \{\hat{j}\}$}}
\textbf{Return $J_l$}
\caption{Computing $J_l$}
\label{alg:demand}
\end{algorithm}

We next discuss the time complexity of Algorithm \ref{alg:KC}. The time complexity of computing $J_l$ as in Algorithm \ref{alg:demand} is $O(|\M|)$ for each $\l \in \L$ (each rider is counted at most once in Algorithm \ref{alg:demand}). Additionally, we know from Sec. \ref{sec:tractable_pooling} that the sum of route capacities given $\kopt$ equals to the maximum capacity of the network $C$. Thus $|\L|=C$, and the time complexity of each iteration of Algorithm \ref{alg:KC} is $O(|\M|C)$. Moreover, riders' utilities are non-decreasing and at least one rider increases their utility by $\epsilon$ in each iteration. Besides, riders' utilities can not exceed the maximum trip value $\Vmax$, because otherwise $J_l=\emptyset$ for all $l \in \L$ regardless of the assigned set $\bl$; thus Algorithm \ref{alg:KC} must terminate before the utility exceeds $\Vmax$. We can conclude that Algorithm \ref{alg:KC} terminates in less than $M \Vmax/\epsilon$ iterations, and its time complexity is $O\(\frac{\Vmax}{\epsilon} |\M|^2 C\)$.





We summarize that $\xopt$ is computed in the following two steps: \\
\emph{Step 1:} Compute the optimal route capacity vector $\kopt$ from Algorithm \ref{alg:flow}. \footnote{This step can be omitted if the network is parallel with vector $\kopt = \(\qr\)_{\r \in \R}$. }\\
\emph{Step 2:} Compute $\yopt$ from Algorithm \ref{alg:KC}. Derive the optimal trip organization vector $\xopt=\chi(\yopt)$.

\vspace{0.2cm}
\noindent\textbf{Computing equilibrium payments $\popt$ and toll prices $\tollopt$.} Given the optimal trip vector $\xopt$, we compute the set of rider payments $\popt$ and toll prices $\tollopt$ such that $\(\xopt, \popt, \tollopt\)$ is a market equilibrium. Recall from Theorem \ref{theorem:primal_dual}, the riders' utilities and toll prices $\(\uopt, \tollopt\)$ in any market equilibrium are optimal solutions of the dual program \eqref{eq:D1bar}. Sec. \ref{sec:tractable_pooling} constructed the augmented trip value function $\W$, which satisfies monotonicity and gross substitutes conditions. 
 Following the same proof ideas as in Theorem \ref{theorem:primal_dual}, we can show that the utility vector $\uopt$ and toll prices $\tollopt$ also can be solved from the following dual program with the augmented trip value function:
\begin{subequations}
        \makeatletter
        \def\@currentlabel{$\mathrm{\overline{D}}$}
        \makeatother
        \label{eq:D1}
        \renewcommand{\theequation}{$\mathrm{\overline{D}}$.\alph{equation}}
\begin{align}
    \min_{\u, \toll} \quad &U(\u, \toll)= \sum_{\m \in \M} \um+ \sum_{\e \in \E} \qe \tolle \notag \\
s.t. \quad & \sum_{\m \in \ball} \um +\sum_{\e \in \r} \tolle \geq \W_r(\ball), \quad \forall \ball \in \Ball, \quad \forall \r \in \R, \label{subeq:D11tilde}\\
        & \um \geq 0, \quad \tolle \geq 0, \quad \forall \m \in \M, \quad \forall \e \in \E. \label{subeq:D12tilde}
\end{align}
\end{subequations}The linear program \eqref{eq:D1} has $|\M|+ |\E|$ number of variables and $|\R| \times |\Ball|$ number of constraints. This linear program can be solved by the ellipsoid method. In each iteration of this method, we need to solve a separation problem to decide whether or not a solution $\(\u, \toll\)$ is feasible, and if not find the constraint that it violates. Since the trip value function $\W$ is monotonic and satisfies the gross substitutes condition, we can solve the separation problem using Algorithm \ref{alg:demand}. For each route $\r \in \R$, we compute $\bar{\b}_r \in \argmax_{\ball \in \Ball} \{\W_r(\ball) - \sum_{\m \in \ball} \um\}$ using Algorithm \ref{alg:demand}. Then, by checking whether or not $\sum_{\m \in \ball_r} \um + \sum_{\e \in \r} \tolle \geq \W_r(\ball_r)$, we can determine if the constraint \eqref{subeq:D11tilde} is satisfied for all route $\r \in \R$. In this way, we solve the separation problem in time polynomial in $|\M|$ and $|\R|$. Thus, the optimal solution of \eqref{eq:D1} can also be solved by ellipsoid method in time polynomial in $|\M|$ and $|\R|$. 

Finally, given any optimal solution $\(\uopt, \tollopt\)$, the riders' payment vector $\popt$ can be obtained from \eqref{eq:u_p}. Thus, we obtain $\(\xopt, \popt, \tollopt\)$ as a market equilibrium. 

Notice that the set of equilibrium utility and toll prices $\(\uopt, \tollopt\)$ may not be singleton. From strong duality theory, we know that the sum of riders' equilibrium utilities and toll prices, must equal to the optimal social welfare given the organized trips in $\xopt$, i.e. $\sum_{\m \in \M} \umopt + \sum_{\e \in \E} \qe \tollopt_e = S(\xopt)$. Therefore, different market equilibria can result in different splits of social welfare between the riders' utilities and the collected toll prices. Next, we highlight a specific market equilibrium that provides the maximum share of social welfare to riders and collects the minimum tolls. 

\section{Strategyproofness and Maximum Rider Utilities}\label{sec:strategyproof}
In this section, we consider the situation where the market is facilitated by a platform that implements a market equilibrium based on the reported preferences of each rider. Two questions arise in this situation: The first is whether or not riders truthfully report their preference parameters to the platform. The second is which market equilibrium is implemented and how it determines the splits between riders' utilities and collected tolls. We show that there exists a strategyproof market equilibrium under which riders truthfully report their preferences. Moreover, this market equilibrium also achieves the maximum utility for all riders and the total toll is the minimum.

We first introduce the definition of strategyproofness. To distinguish between the true preference parameters and the reported preference parameters, we denote the reported parameters as $\alpha^{'}$ and $\beta^{'}$.\footnote{We assume that riders have homogeneous carpool disutility that is known by the platform.} The corresponding market equilibrium is denoted $\(x^{*'}, p^{*'}, \toll^{*'}\)$. The utility vector under market equilibrium with the true preference parameters (resp. reported preference parameters) $\uopt$ (resp. $u^{*'}$) can be computed as in \eqref{eq:u_p}.  
\begin{definition}[Strategyproofness]\label{def:strategyproof}
A market equilibrium $\(\xopt, \popt, \tollopt\)$ is strategyproof if for any preference parameters $\alpha^{'} \neq \alpha$ and $\beta^{'} \neq \beta$, $\umopt \geq u^{m*'}$ for all $\m \in \M$.   
\end{definition}


We next define the Vickery-Clark-Grove (VCG) Payment vector. For each $\m \in \M$, we denoted $\xoptmm$ as the optimal trip vector when rider $\m$ is not present. The social welfare for riders in $\M \setminus \{\m\}$ given the optimal trip vector $\xoptmm$ is denoted $S^{-\m}(\xoptmm) = \sum_{\b \in \B} \sum_{\r \in \R} \Vrb \xoptmm_{\r}(\b)$ , and the social welfare for riders in $\M \setminus \{\m\}$ with $\xopt$ is $S_{-\m}(\xopt)= S(\xopt) - \sum_{\b\ni \m}\sum_{\r \in \R} \vrmb \xrbopt$. 
\begin{definition} A VCG payment vector
$\pdag= \(\pmdag\)_{\m \in \M}$ is given by: 
\begin{align}\label{eq:pmdag}
    \pmdag 
    = S_{-\m}(\xoptmm) - S_{-\m}(\xopt), \quad \forall \m \in \M. 
\end{align}
\end{definition}

In VCG payment vector \eqref{eq:pmdag}, each rider $\m$'s payment is the difference of the total trip values for all other riders with and without rider $\m$, i.e. $\pmdag$ is the externality of each rider $\m$ on all other riders. Under the optimal trip vector $\xopt$ and the VCG payment vector $\pdag$, the utility vector $\udag = \(\udagm\)_{\m \in \M}$ is given by: \begin{align}\label{eq:umaxm}
    \umaxm &\stackrel{\eqref{eq:u_p}}{=} \sum_{\b\ni \m}\sum_{\r \in \R} \Vrb \xrbopt - \pmdag \stackrel{\eqref{eq:pmdag}}{=}  
    S(\xopt) - S_{-\m} (\xopt_{-m}), \quad \forall \m \in \M. 
\end{align}
That is, the utility of each rider $\m \in \M$ is the difference of the optimal social welfare with and without rider $\m$. 
\begin{lemma}[\cite{vickrey1961counterspeculation}]\label{lemma:VCG}
A market equilibrium is strategyproof if the payment vector is $\pdag$. 
\end{lemma}

The next theorem shows that there exists a toll price vector such that the market equilibrium payment vector is $\pdag$ and the riders' utility vector is $\udag$. This market equilibrium is strategyproof. Moreover, all riders' utilities are higher than that under any other market equilibrium, and the total collected tolls is the minimum. 

\begin{theorem}\label{theorem:strategyproof}
There exists a toll price vector $\tolldag$ such that $\(\xopt, \pdag, \tolldag\)$ is a market equilibrium, and is strategyproof. Moreover, for any other market equilibrium $\(\xopt, \popt, \tollopt\)$, 
\begin{align*}
    \umaxm &\geq \umopt, \quad \forall \m \in \M, \quad 
    \sum_{\e \in \E} \qe \tolldag_e \leq \sum_{\e \in \E} \qe \tollopt_e. 
\end{align*}
\end{theorem}

We denote the set of $\uopt$ in the optimal solutions of the dual problem \eqref{eq:D1bar} as $\Uopt$. From Theorem \ref{theorem:primal_dual}, we know that any utility vector $\uopt$ is an equilibrium utility vector if and only if there exists a toll price vector $\tollopt$ such that $\(\uopt, \tollopt\)$ is an optimal solution of \eqref{eq:D1bar}, i.e. $\uopt \in \Uopt$. To show that $\udag$ is the maximum equilibrium utility vector, we need to prove that $\udag$ is the maximum component in the set $\Uopt$.

We proceed in three steps: Firstly, Lemma \ref{lemma:utility} shows that the set $\Uopt$ is equivalent to the set of utility vectors in the optimal solution set of the dual program of \eqref{eq:LP2k}. Secondly, the set of optimal utility vectors in the dual program of \eqref{eq:LP2k} is the same as the set of prices in Walrasian equilibrium of the equivalent economy constructed in Sec. \ref{sec:tractable_pooling} (Lemma \ref{lemma:utility_walrasian}). Finally, the set of good prices in Walrasian equilibrium is a complete lattice, and the maximum component is $\udag$ as in \eqref{eq:umaxm} (Lemma \ref{lemma:lattice}). 


We now present the formal statements of these lemmas and their proof ideas. The proofs are included in Appendix B. 

\begin{lemma}\label{lemma:utility}
A utility vector $\uopt \in \Uopt$ if and only if there exists vector $\lambda^* = \(\lambda^*_r\)_{\r \in \R}$ such that $\(\uopt, \lambda^*\)$ is an optimal solution of the following linear program:
\begin{subequations}
        \makeatletter
        \def\@currentlabel{D$k^*$}
        \makeatother
        \label{eq:D2k}
        \renewcommand{\theequation}{D$k^*$.\alph{equation}}
    \begin{align}
         \min_{\u, \lambda}  \quad & \sum_{\m \in \M} \um + \sum_{\r \in \Ropt} \kopt_r \lambdar, \notag\\
   s.t. \quad  &  \sum_{\m \in \b} \um + \lambdar \geq \Vrb \quad \forall \r \in \Ropt, \quad \forall \b \in \B,  \label{subeq:D2k1}\\
    &\um \geq 0, \lambdar \geq 0, \quad \forall \m \in \M, \quad \forall \r \in \Ropt,\label{subeq:D2k2}
    \end{align}
\end{subequations}
where $\lambdar$ is the dual variable of constraint \eqref{subeq:LP2k2} for each $\r \in \R$. 
\end{lemma}


In \eqref{eq:D2k}, the dual variable $\lambdar$ can be viewed as the toll price set on each route $\r \in \Ropt$. We note that \eqref{eq:D2k} is less restrictive than \eqref{eq:D1bar}, which is the dual program on the original network, in two respects: Firstly, constraints \eqref{eq:D2k} are only set for the set of routes $\Ropt$ of the sub-network rather than on all routes in the whole network. Secondly, the toll prices $\lambda$ in \eqref{eq:D2k} are set on routes instead of on edges as in $\toll$ of \eqref{eq:D1bar}. Any edge toll price vector can be equivalently represented as toll prices on routes by summing the tolls of all edges on any route. Therefore, given any feasible solution $\(\u, \toll\)$ of \eqref{eq:D1bar}, $\(\u, \lambda\)$ where $\lambda_r= \sum_{\e \in \r}\toll_e$ for each $\r \in \Ropt$ is also feasible in \eqref{eq:D2k}.

We can check that for any optimal solution $\(\uopt, \tollopt\)$ of \eqref{eq:D1bar}, the vector $\(\uopt, \lambda^*\)$ -- where $\lambda^*_r = \sum_{\e \in \r} \tollopt_e$ for each $\r \in \Ropt$ -- must also be optimal in \eqref{eq:D2k}. That is, the set $\Uopt$ is a subset of the optimal utility vectors in \eqref{eq:D2k}. This result follows from strong duality theory and Lemma \ref{lemma:FF}: From the strong duality theory, the optimal values of the objective function in \eqref{eq:D2k} (resp. \eqref{eq:LP1bar}) equals to the optimal value of the primal problems \eqref{eq:LP1bar} (resp. \eqref{eq:LP2k}). From Lemma \ref{lemma:FF}, we know that the optimal trip organization vector is the same in both \eqref{eq:LP1bar} and \eqref{eq:LP2k}. Thus, the optimal value of \eqref{eq:D1bar} is the same as that of \eqref{eq:LP2k}. Since the value of the objective function with $\(\uopt, \tollopt\)$ equals to that with $\(\uopt, \lambda^*\)$, we know that $\(\uopt, \lambda^*\)$ must be an optimal solution of \eqref{eq:D2k}. 

Furthermore, we can show that for any optimal solution $\uopt$ of \eqref{eq:D2k}, there must exist an edge toll vector $\tollopt$ such that $\(\uopt, \tollopt\)$ is an optimal solution of \eqref{eq:D1}. That is, any equilibrium utility vector with route toll prices on the sub-network can also be induced by edge toll prices on the original network. This result relies on the fact that the network is series parallel, and it is proved by mathematical induction. 

Lemma \ref{lemma:utility} enables us to characterize the riders' utility set $\Uopt$ using the less restrictive dual program \eqref{eq:D2k}. Recall that in Sec. \ref{sec:tractable_pooling}, we have shown that the trip organization problem on the constructed augmented network with the augmented value function is equivalent to an economy with indivisible goods (Lemma \ref{lemma:equivalent_economy}). The next lemma shows that the set $\Uopt$ is the same as the set of Walrasian equilibrium prices in the equivalent economy. 
\begin{lemma}\label{lemma:utility_walrasian}
A utility vector $\uopt \in \Uopt$ if and only if there exists $\yall^*$ such that $\(\yall^*, \uopt\)$ is a Walrasian equilibrium of the economy.
\end{lemma}

Moreover, since the augmented trip value function $W$ is monotonic and satisfies gross substitutes condition, the set of Walrasian equilibrium price vectors is a lattice, and has a maximum component. 

\begin{lemma}[\cite{gul1999walrasian}]\label{lemma:lattice}
If the value function $W$ satisfies the monotonicity and gross substitutes conditions, then the set of Walrasian equilibrium prices is a lattice and has a maximum component $\umax = \(\umaxm\)_{\m \in \M}$ as in \eqref{eq:umaxm}. 
\end{lemma}

From Lemmas \ref{lemma:utility} -- \ref{lemma:lattice}, we know that $\udag$ is the maximum component in the set $\Uopt$. That is, there exists a toll price vector $\tolldag = \(\tolldag_{e}\)_{\e \in \E}$ such that $\(\udag, \tolldag\)$ is an optimal solution of \eqref{eq:D1bar}, and hence $\(\xopt, \pdag, \tolldag\)$ is a market equilibrium. Additionally, from Lemma \ref{lemma:VCG}, we know that this market equilibrium is strategyproof. Moreover, all riders achieve the maximum equilibrium utilities in the equilibrium. Since $\sum_{\m \in \M} \umopt + \sum_{\e \in \E} \qe \tollopt_e= S(\xopt)$ for any market equilibrium $\(\xopt, \uopt, \tollopt\)$, this also implies that the total amount of tolls $\sum_{\e \in \E} \qe \tolldag_e$ that is collected in market equilibrium $\(\xopt, \pdag, \tolldag\)$ is the minimum. We thus conclude Theorem \ref{theorem:strategyproof}.

Finally, we discuss the computation of the market equilibrium $\(\xopt, \pdag, \tolldag\)$. In particular, the optimal trip assignment $\xopt$ can be computed in two steps described in Sec. \ref{sec:algorithm} using Algorithms \ref{alg:flow} -- \ref{alg:KC}. Then, we re-run Algorithm \ref{alg:KC} given $\kopt$ and rider set $\M \setminus \{\m\}$ to compute $\xoptmm$ for each $\m \in \M$. We compute the utility vector $\udag$ (resp. payment vector $\pdag$) as in \eqref{eq:umaxm} (resp. \eqref{eq:pmdag}). 

For any $\e \in \E$, we set $\tolldag_e=0$ if $\sum_{\b \in \B} \sum_{\r \ni \e}\xrbopt < \qe$. From \eqref{eq:D1}, we know that $\tolldag$ is any vector that satisfies the following constraints: 
\begin{equation}\label{eq:tolldag}
\begin{split}
    \sum_{\e \in \r} \tolldag_e &= \max_{\ball \in \Ball} \W_r(\ball) - \sum_{\m \in \ball} \umaxm, \quad \forall \r \in \Ropt, \\
    \sum_{\e \in \r} \tolldag_e & \geq \max_{\ball \in \Ball} \W_r(\ball) - \sum_{\m \in \ball} \umaxm, \quad \forall \r \in \R\setminus \Ropt. 
\end{split}
\end{equation}
Finding a vector $\tolldag$ that satisfies constraints in \eqref{eq:tolldag} is equivalent to solving a linear program with a constant objective function and feasibility constraints \eqref{eq:tolldag}. This linear program can be computed by the ellipsoid method, in which the separation problem in each iteration is to check whether or not the toll price vector $\tolldag$ satisfies the feasible constraints in \eqref{eq:tolldag}. Since the augmented trip value function $\W$ satisfies monotonicity and gross substitutes condition, we can compute the right-hand-side value of the constraint in \eqref{eq:tolldag} using Algorithm \ref{alg:demand} in time $O(|\M|)$ for each $\r \in \R$. That is, the separation problem in each iteration can be computed in polynomial time of $|\M|$ and $|\R|$. Therefore, a toll vector $\tolldag$ that satisfies \eqref{eq:tolldag} can be computed in polynomial time of $|\M|$ and $|\R|$.

 \section{Concluding Remarks}
In this article, we studied the existence and computation of market equilibrium for organizing socially efficient carpooled trips over a transportation network using autonomous cars. We also identified a market equilibrium that is strategyproof and maximizes riders' utilities. Our approach can be used to analyze incentive mechanisms for sharing limited resources in networked environment. 
 
 One interesting direction for future work is to characterize equilibrium in a transportation market when riders belong to different classes that are differentiated by their carpool disutility levels. In this situation, riders with different carpool disutilies may be grouped into trips that are organized using different vehicle sizes to reflect the riders' car sharing preferences. 
 
 A more general problem is to design market with both autonomous and human-driven carpooled trips, wherein riders may have different preferences of over these service types. A pre-requisite to the design of such a market is quantitative evaluation of how autonomous and human-driven vehicles differ in terms of their utilization of road capacity and the incurred route travel times \cite{jin2020analysis}. Analysis of differentiated pricing and tolling schemes corresponding to trip assignments between the two service types is an interesting and relevant problem for future work.

\bibliographystyle{plain}
\bibliography{library.bib}

\begin{thebibliography}{10}

\bibitem{afeche2018ride}
Philipp Afeche, Zhe Liu, and Costis Maglaras.
\newblock Ride-hailing networks with strategic drivers: The impact of platform
  control capabilities on performance.
\newblock {\em Columbia Business School Research Paper}, (18-19):18--19, 2018.

\bibitem{ashlagi2019edge}
Itai Ashlagi, Maximilien Burq, Chinmoy Dutta, Patrick Jaillet, Amin Saberi, and
  Chris Sholley.
\newblock Edge weighted online windowed matching.
\newblock In {\em Proceedings of the 2019 ACM Conference on Economics and
  Computation}, pages 729--742, 2019.

\bibitem{banerjee2015pricing}
Siddhartha Banerjee, Ramesh Johari, and Carlos Riquelme.
\newblock Pricing in ride-sharing platforms: A queueing-theoretic approach.
\newblock In {\em Proceedings of the Sixteenth ACM Conference on Economics and
  Computation}, pages 639--639, 2015.

\bibitem{banerjee2018state}
Siddhartha Banerjee, Yash Kanoria, and Pengyu Qian.
\newblock State dependent control of closed queueing networks with application
  to ride-hailing.
\newblock {\em arXiv preprint arXiv:1803.04959}, 2018.

\bibitem{bein1985minimum}
Wolfgang~W Bein, Peter Brucker, and Arie Tamir.
\newblock Minimum cost flow algorithms for series-parallel networks.
\newblock {\em Discrete Applied Mathematics}, 10(2):117--124, 1985.

\bibitem{bikhchandani1997competitive}
Sushil Bikhchandani and John~W Mamer.
\newblock Competitive equilibrium in an exchange economy with indivisibilities.
\newblock {\em Journal of Economic Theory}, 74(2):385--413, 1997.

\bibitem{bimpikis2019spatial}
Kostas Bimpikis, Ozan Candogan, and Daniela Saban.
\newblock Spatial pricing in ride-sharing networks.
\newblock {\em Operations Research}, 67(3):744--769, 2019.

\bibitem{cachon2017role}
Gerard~P Cachon, Kaitlin~M Daniels, and Ruben Lobel.
\newblock The role of surge pricing on a service platform with self-scheduling
  capacity.
\newblock {\em Manufacturing \& Service Operations Management}, 19(3):368--384,
  2017.

\bibitem{castillo2017surge}
Juan~Camilo Castillo, Dan Knoepfle, and Glen Weyl.
\newblock Surge pricing solves the wild goose chase.
\newblock In {\em Proceedings of the 2017 ACM Conference on Economics and
  Computation}, pages 241--242, 2017.

\bibitem{de2003combinatorial}
Sven De~Vries and Rakesh~V Vohra.
\newblock Combinatorial auctions: A survey.
\newblock {\em INFORMS Journal on computing}, 15(3):284--309, 2003.

\bibitem{gul1999walrasian}
Faruk Gul and Ennio Stacchetti.
\newblock Walrasian equilibrium with gross substitutes.
\newblock {\em Journal of Economic theory}, 87(1):95--124, 1999.

\bibitem{gurvich2015dynamic}
Itai Gurvich and Amy Ward.
\newblock On the dynamic control of matching queues.
\newblock {\em Stochastic Systems}, 4(2):479--523, 2015.

\bibitem{jin2020analysis}
Li~Jin, Mladen Cicic, Karl~H Johansson, and Saurabh Amin.
\newblock Analysis and design of vehicle platooning operations on mixed-traffic
  highways.
\newblock {\em IEEE Transactions on Automatic Control}, 2020.

\bibitem{kelso1982job}
Alexander~S Kelso~Jr and Vincent~P Crawford.
\newblock Job matching, coalition formation, and gross substitutes.
\newblock {\em Econometrica: Journal of the Econometric Society}, pages
  1483--1504, 1982.

\bibitem{leme2017gross}
Renato~Paes Leme.
\newblock Gross substitutability: An algorithmic survey.
\newblock {\em Games and Economic Behavior}, 106:294--316, 2017.

\bibitem{lian2020autonomous}
Zhen Lian and Garrett van Ryzin.
\newblock Autonomous vehicle market design.
\newblock {\em Available at SSRN}, 2020.

\bibitem{milchtaich2006network}
Igal Milchtaich.
\newblock Network topology and the efficiency of equilibrium.
\newblock {\em Games and Economic Behavior}, 57(2):321--346, 2006.

\bibitem{ostrovsky2019carpooling}
Michael Ostrovsky and Michael Schwarz.
\newblock Carpooling and the economics of self-driving cars.
\newblock In {\em Proceedings of the 2019 ACM Conference on Economics and
  Computation}, pages 581--582, 2019.

\bibitem{ozkan2020dynamic}
Erhun {\"O}zkan and Amy~R Ward.
\newblock Dynamic matching for real-time ride sharing.
\newblock {\em Stochastic Systems}, 10(1):29--70, 2020.

\bibitem{reijnierse2002verifying}
Hans Reijnierse, Anita van Gellekom, and Jos~AM Potters.
\newblock Verifying gross substitutability.
\newblock {\em Economic Theory}, 20(4):767--776, 2002.

\bibitem{siddiq2019ride}
Auyon Siddiq and Terry Taylor.
\newblock Ride-hailing platforms: Competition and autonomous vehicles.
\newblock {\em Available at SSRN 3426988}, 2019.

\bibitem{vickrey1961counterspeculation}
William Vickrey.
\newblock Counterspeculation, auctions, and competitive sealed tenders.
\newblock {\em The Journal of finance}, 16(1):8--37, 1961.

\end{thebibliography}

\newpage
\begin{appendix}
\section{Proof of Section \ref{sec:general}}\label{apx:proof_A}

\noindent\emph{Proof of Theorem \ref{theorem:primal_dual}.} 
First, we proof that the four conditions of market equilibrium $\(\xopt, \popt, \tollopt\)$ ensures that $\xopt$ satisfies the feasibility constraints of the primal \eqref{eq:LP1bar}, $\(\uopt, \tollopt\)$ satisfies the dual \eqref{eq:D1bar}, and $\(\xopt, \uopt, \tollopt\)$ satisfies the complementary slackness conditions. The vector $\uopt$ is the utility vector computed from \eqref{eq:u_p}.  
\begin{enumerate}
\item[(i)] Feasibility constraints of \eqref{eq:LP1bar}. Since $\xbaropt$ is a feasible trip vector, $\xbaropt$ must satisfy the Feasibility constraints of \eqref{eq:LP1bar}. 

\item[(ii)] Feasibility constraints of \eqref{eq:D1bar}. From the stability condition \eqref{eq:stability}, individual rationality \eqref{eq:ir}, and the fact that toll prices are non-negative, we know that $\(\uopt, \tollopt\)$ satisfies the feasibility constraints of \eqref{eq:D1bar}. 

\item[(iii)] Complementary slackness condition with respect to  \eqref{subeq:LP11}. If rider $\m$ is not assigned, then \eqref{subeq:LP11} is slack with the integer trip assignment $\xbaropt$ for some rider $\m$. The budget balanced condition \eqref{subeq:p_not_assigned} shows that $\popt_m=0$. Since rider $\m$ is not in any trip and the payment is zero, the dual variable (i.e. rider $\m$'s utility) $\umopt=0$. On the other hand, if $\umopt>0$, then rider $\m$ must be in a trip, and constraint \eqref{subeq:LP11} must be tight. Thus, we can conclude that the complementary slackness condition with respect to the primal constraint \eqref{subeq:LP11} is satisfied. 

\item[(iv)] Complementary slackness condition with respect to \eqref{subeq:LP12}. Since the mechanism is market clearing, toll price $\tolle$ is nonzero if and only if the load on edge $\e$ is below the capacity, i.e. the primal constraint \eqref{subeq:LP12} is slack for edge $\e\in \E$. Therefore, the complementary slackness condition with respect to the primal constraint \eqref{subeq:LP12} is satisfied. 
\item[(v)] Complementary slackness condition with respect to \eqref{subeq:D11}. From \eqref{subeq:bb}, we know that for any organized trip, the corresponding dual constraint \eqref{subeq:D11} is tight. If constraint \eqref{subeq:D11} is slack for a trip $\(\bbar, \r\)$, then the budget balance constraint ensures that trip is not organized. Therefore, the complementary slackness condition with respect to the primal constraint \eqref{subeq:D11} is satisfied. 
\end{enumerate}

We can analogously show that the inverse of (i) -- (v) are also true: the feasibility constraints of \eqref{eq:LP1bar} and \eqref{eq:D1bar}, and the complementary slackness conditions ensure that $\(\xopt, \popt, \tollopt\)$ is a market equilibrium. Thus, we can conclude that $\(\xopt, \popt, \tollopt\)$ is a market equilibrium if and only if $\(\xopt, \uopt, \tollopt\)$ satisfies the feasibility constraints of \eqref{eq:LP1bar} and \eqref{eq:D1bar}, and the complementary slackness conditions.

From strong duality theory, we know that the equilibrium trip vector $\xbaropt$ must be an optimal integer solution of \eqref{eq:LP1bar}. Therefore, the existence of market equilibrium is equivalent to the existence of an integer optimal solution of \eqref{eq:LP1bar}. The optimal trip assignment is an optimal integer solution of \eqref{eq:LP1bar}, and $\(\uopt, \tollopt\)$ is an optimal solution of the dual problem \eqref{eq:D1bar}. The payment $\popt$ can be computed from \eqref{eq:u_p}. \QEDA

\vspace{0.2cm}
\noindent\emph{Proof of Corollary \ref{cor:shorter}.} Consider any two routes $\r, \r' \in \R$ such that $\tr \geq t_{\r'}$. Given $\xopt$, we denote the rider group that takes route $\r$ as $\bbar_r$. If no rider group is assigned to route $\r$, then we denote $\bbar_r = \emptyset$. From \eqref{subeq:bb}, we have
\begin{align*}
    \sum_{\m \in \bbar_r} \uopt_m + \sum_{\e \in \r} \tollopt_e = V_r(\bbar_r).
\end{align*}
Additionally, since $\(\uopt, \tollopt\)$ satisfies constraint \eqref{subeq:D11}, we know that 
\begin{align*}
     \sum_{\m \in \bbar_r} \uopt_m + \sum_{\e \in \r'} \tollopt_e \geq V_r'(\bbar_r). 
\end{align*}
Therefore, we must have: 
\begin{align*}
    &\sum_{\e \in \r'} \tollopt_e - \sum_{\e \in \r} \tollopt_e \geq \Vbar_r'(\bbar_r) - \Vbar_r(\bbar_r) = \(\sum_{\m \in \bbar_r} \tripm - \sum_{\m \in \bbar_r} \vm t_{\r'} - \sum_{\m \in \bbar_r} \gamma^m(|\bbar_r|) t_{\r'} - \delta|\bbar_r| t_{\r'}\)\\
    &- \(\sum_{\m \in \bbar_r} \tripm - \sum_{\m \in \bbar_r} \vm  \tr - \sum_{\m \in \bbar_r} \gamma^m(|\bbar_r|) \tr - \delta|\bbar_r|\tr\)\\
    =& \(\sum_{\m \in \bbar_r} \vm+ \sum_{\m \in \bbar_r} \gamma^m(|\bbar_r|) + \delta|\bbar_r|\)\(\tr- t_{\r'}\)\geq 0.
\end{align*}\QEDA

\section{Proof of Section \ref{sec:tractable_pooling}.}\label{apx:proof_B}
\vspace{0.3cm}
\noindent\emph{Proof of Lemma \ref{lemma:FF}.}
Consider any (fractional) optimal solution of \eqref{eq:LP1bar}, denoted as $\xhat$. We denote $\fhat(\b) = \sum_{\r \in \R} \xhat_r(\b)$ as the flow of group $\b$, and $\widehat{F}= \sum_{\b \in \B} \fhat(\b)$ is the total flows. Since $\xhat$ is feasible, we know that $\widehat{F} \leq C$, where $C$ is the maximum capacity of the network. For each $\b \in \B$, we re-write the trip valuation as follows: 
\begin{align*}
V_r(\b)= z(\b)- g(\b) \tr, \quad \forall \(\b, \r\) \in \B \times \R, \end{align*}
where $g(\b) = \sum_{\m \in \b} \vm + \sum_{\m \in \b} \gamma(|\b|)+  \unitcost|\b|$, and $z(\b)= \sum_{\m \in \b} \tripm$.

The set of all groups with positive flow in $\xhat$ is $\widehat{\B} \deleq \{\bhat \in \B|\fhat(\bhat)>0\}$. We denote the number of rider groups in $\Bhat$ as $n$, and re-number these rider groups in decreasing order of $g(\bhat)$, i.e. 
\begin{align*}
g(\bhat_1) \geq g(\bhat_2) \geq \cdots \geq g(\bhat_n).
\end{align*}

We now construct another trip  vector $\xopt$ by the following procedure: \\
\emph{Initialization:} Set route set $\tildeR = \Ropt$, route capacity $\tildeq_r = \kopt_{\r}$ for all $ \forall \r \in \tildeR$, and initial zero assignment vector $\xrbopt \leftarrow 0$ for all $\r \in \R$ and all $\b \in \B$\\
\emph{For $j=1, \dots, n$:}
\begin{itemize}
    \item[(a)] Assign rider group $\bhat_j$ to a route 
    $\rhat$ in $\tildeR$, which has the minimum travel time among all routes with flow less than the capacity, i.e. $\rhat \in \argmin_{\r \in \{\tildeR| \sum_{\b \in \B} \xrbopt < \tildeq_r\}}\{\tr\}$. 
    \item[(b)] If $\sum_{\b \in \B} \xrbopt + \fhat(\bhat_j) \leq \tildeq_r$, then $\xopt_r(\bhat_j) = \fhat(\bhat_j)$. 
    \item[(c)] Otherwise, assign $\xopt_r(\bhat_j)= \tildeq_r - \sum_{\b \in \B} \xrbopt$, and continue to assign the remaining weight of rider group $\bhat_j$ to the next unsaturated route with the minimum cost. Repeat this process until the condition in (b) is satisfied, i.e. the total weight $\fhat(\b_j)$ is assigned. 
\end{itemize}

We can check that $\sum_{\b \ni \m} \sum_{r \in \R}  \xrbopt = \sum_{\b \ni \m} \fhat(\b) \leq 1$ so that \eqref{subeq:LP2k1} is satisfied. Additionally, since in the assignment procedure, the total weight assigned to route $\r$ is less than or equal to $\kopt_r$, we must have $\sum_{\b \in \B} \xrbopt \leq \kopt_r$ for all $\r \in \R$, i.e. \eqref{subeq:LP2k2} is satisfied. Thus, $\xopt$ is a feasible solution of \eqref{eq:LP2k}. 

It remains to prove that $\xopt$ is optimal of \eqref{eq:LP2k}. We prove this by showing that $V(\xopt) \geq V(\xhat)$. The objective function $S(\xopt)$ can be written as follows:
\begin{align}\label{eq:decompose_V}
&\sum_{r \in \R} \sum_{\b \in \B} \Vrb \xrbopt = \sum_{r \in \R} \sum_{\b \in \B} z(\b) \xrbopt - \sum_{r \in \R} \sum_{\b \in \B} g(\b) \tr \xrbopt. \end{align}
We note that since $\sum_{\r \in \R} \kopt_r = C$ and $\sum_{\r \in \R} \sum_{\b \in \B} \xhat_r(\b) \leq C$, the algorithm must terminate with all groups in $\xhat$ being assigned. Therefore, $\sum_{\r \in \R} \xrbopt=\fhat(\b) = \sum_{\r \in \R} \xhat_r(\b)$ for all $\b \in \B$. Therefore, 
\begin{align}\label{eq:equal_part}
&\sum_{r \in \R} \sum_{\b \in \B} z(\b) \xrbopt = \sum_{\b \in \B} z(\b) \fhat(\b) = \sum_{r \in \R} \sum_{\b \in \B} z(\b) \xhat_r(\b)
\end{align}
Then, $V(\xopt) \geq V(\xhat)$ is equivalent to $\sum_{r \in \R}  \sum_{\b \in \B} g(\b)\tr \xrbopt \leq \sum_{r \in \R}  \sum_{\b \in \B} g(\b)\tr \xhat_r^*$. To prove this, we show that $\xopt$ minimizes the term $\sum_{r \in \R} \sum_{\b \in \B} g(\b) \tr \xrbopt$ among all feasible $\x$ that induces the same flow of groups as $\xhat$, i.e. 
\begin{align}\label{eq:induction}
\xopt \in \argmin_{\x \in \X(\fhat)}\sum_{r \in \R}  \sum_{\b \in \B} g(\b)\tr \xrb, 
\end{align}
where 
\begin{align}\label{eq:X}
\X(\fhat) \deleq \left\{\(\xrb\)_{\r \in \R, \b \in \B}\left\vert
\begin{array}{l}
\sum_{\r \in \R} \xrb= \fhat(\b), \quad \forall \b \in \B, \\
\sum_{\b \in \B}\sum_{\r \ni \e} \xrb \leq \qe, \quad \forall \e \in \E, \\
\xrb \geq 0, \quad \forall \r \in \R, \quad \forall \b \in \B. 
\end{array}
\right.
\right\}
\end{align}

We prove \eqref{eq:induction} by mathematical induction. To begin with, \eqref{eq:induction} holds trivially on any single-link network. We ext prove that if \eqref{eq:induction} holds on two series-parallel sub-networks $\Gp$ and $\Gpp$, then \eqref{eq:induction} holds on the network $\G$ that connects $\Gp$ and $\Gpp$ in series or in parallel. In particular, we analyze the cases of series connection and parallel connection separately: 

\vspace{0.2cm}
\noindent\emph{(Case 1)} Series-parallel Network $G$ is formed by connecting two series-parallel sub-networks $\Gp$ and $\Gpp$ in series. \\
    We denote the set of routes in subnetwork $\Gp$ and $\Gpp$ as $\Rp$ and $\Rpp$, respectively. Since $\Gp$ and $\Gpp$ are connected in series, the set of routes in network $\G$ is $\R \deleq \Rp \times \Rpp$. For any flow vector $\fhat$, we define the set of trip vectors on $G$ that satisfy the constraint in \eqref{eq:induction} as $\X(\fhat)$. We also define the trip  vector that is obtained from the above-mentioned procedure based on $\fhat$ as $\xopt$.

    Since the two sub-networks are connected in sequence, the group flow vectors in $\Gp$ and $\Gpp$ are also $\fhat$. Analogously, we define the set of trip vectors on sub-network $\Gp$ (resp. $\Gpp$) that satisfies the constraint in \eqref{eq:induction} as $\Xp(\fhat)$ (resp. $\Xpp(\fhat)$). We can check that $\Xp(\fhat)$ (resp. $\Xpp(\fhat)$) is the set of trip vectors in $\X(\fhat)$ that is restricted on network $\Gp$ (resp. $\Gpp$). That is, for any $\x \in \X(\fhat)$, we can find $\xp \in \Xp(\fhat)$ (resp. $\xpp \in \Xpp(\fhat)$) such that $\sum_{\rpp \in \Rpp}\x_{\rp\rpp}(\b)=\xp_{\rp}(\b)$ (resp. $\sum_{\rp \in \Rp}\x_{\rp\rpp}(\b)=\xpp_{\rpp}(\b)$) for all $\b \in \B$ and all $\rp \in \Rp$ (resp. $\rpp \in \Rpp$).  Since the two subnetworks are connected sequentially, we have the follows: 
    \begin{align}
        &\sum_{r \in \R}  \sum_{\b \in \B} g(\b)\tr \xrb = \sum_{\rp \in \Rp} \sum_{\b \in \B} g(\b)t_{\rp} \(\sum_{\rpp \in \Rpp} \x_{\rp\rpp}(\b)\)+ \sum_{\rpp \in \Rpp} \sum_{\b \in \B} g(\b)t_{\rpp} \(\sum_{\rp \in \Rp} \x_{\rp\rpp}(\b)\)\notag \\
        =&\sum_{\rp \in \Rp} \sum_{\b \in \B} g(\b)t_{\rp} \xp_{\rp}(\b)+ \sum_{\rpp \in \Rpp} \sum_{\b \in \B} g(\b)t_{\rpp} \xpp_{\rpp}(\b).\label{eq:series_1}
        \end{align}

We also denote the trip vector that is obtained from the above-mentioned procedure based on $\fhat$ in $\Gp$ (resp. $\Gpp$) as $\xoptp$ (resp. $\xoptpp$). We now argue that $\sum_{\rpp \in \Rpp}\xopt_{\rp\rpp}(\b)=\xoptp_{\rp}(\b)$ for all $\b \in \B$ and all $\rp \in \Rp$. For the sake of contradiction, assume that there exists $\b \in \B$ such that $\sum_{\rpp \in \Rpp}\xopt_{\rp\rpp}(\b) \neq \xoptp_{\rp}(\b)$ for at least one $\rp \in \Rp$. We denote $\bhat$ as one such group with the maximum $g(\bhat)$. Since the total flow of $\bhat$ is $\fhat(\bhat)$ in both $\xopt$ and $\xoptp$, if $\sum_{\rpp \in \Rpp}\xopt_{\rp\rpp}(\bhat) \neq \xoptp_{\rp}(\bhat)$ on one $\rp \in \Rp$, the same inequality must hold for another $\rptwo \in \Rp$. Without loss of generality, we assume that $t_{\rp}< t_{\rptwo}$. Since any group $\b$ that are assigned before $\bhat$ ($g(\b)< g(\bhat)$) satisfy $\sum_{\rpp \in \Rpp}\xopt_{\rp\rpp}(\b)=\xoptp_{\rp}(\b)$ for all $\rp \in \Rp$, we know that the available route capacities $\tildef$ in the round of assigning $\bhat$ in procedure (i) -- (iii) satisfy $\sum_{\rpp \in \Rpp}\tildef_{\rp\rpp}=\tildef_{\rp}$ for all $\rp \in \Rp$. Therefore, if $\sum_{\rpp \in \Rpp}\xopt_{\rp\rpp}(\bhat)<\xoptp_{\rp}(\bhat)$, then $\xoptp$ is not obtained by procedure (i) -- (iii) on $\Gp$ because $\rp$ is not saturated with $\xoptp$ in the round of assigning $\bhat$, and more flow of $\bhat$ should be moved from $\rptwo$ to $\rp$ to saturate route $\rp$. We can analogously argue that if $\sum_{\rpp \in \Rpp}\xopt_{\rp\rpp}(\bhat)>\xoptp_{\rp}(\bhat)$, then $\xopt$ is not obtained from the algorithm for $G$. In either case, we have arrived at a contradiction. We can analogously argue that $\sum_{\rp \in \Rp}\xopt_{\rp\rpp}(\b)=\xoptpp_{\rpp}(\b)$ for all $\b \in \B$ and all $\rpp \in \Rpp$. Therefore, \begin{align}
        &\sum_{r \in \R}  \sum_{\b \in \B} g(\b)\tr \xrbopt = \sum_{\rp \in \Rp} \sum_{\b \in \B} g(\b)t_{\rp} \(\sum_{\rpp \in \Rpp} \xopt_{\rp\rpp}(\b)\)+ \sum_{\rpp \in \Rpp} \sum_{\b \in \B} g(\b)t_{\rpp} \(\sum_{\rp \in \Rp} \xopt_{\rp\rpp}(\b)\)\notag\\
        =&\sum_{\rp \in \Rp} \sum_{\b \in \B} g(\b)t_{\rp} \xoptp_{\rp}(\b)+ \sum_{\rpp \in \Rpp} \sum_{\b \in \B} g(\b)t_{\rpp} \xoptpp_{\rpp}(\b)\label{eq:series_2}
        \end{align}
        
If \eqref{eq:induction} holds on both sub-networks (i.e. $\xoptp \in \arg\min_{\x \in \Xp(\fhat)}\sum_{\rp \in \Rp}  \sum_{\b \in \B} g(\b)\trp \xp_{\rp}(\b)$ and $\xoptpp \in \arg\min_{\x \in \Xpp(\fhat)}\sum_{\rpp \in \Rpp}  \sum_{\b \in \B} g(\b)\trpp \xpp_{\rpp}(\b)$), then from \eqref{eq:series_1} -- \eqref{eq:series_2}, we know that \eqref{eq:induction} also holds in network $\G$. 
\vspace{0.2cm}
    
\noindent\emph{(Case 2)} Series-parallel Network $G$ is formed by connecting two series-parallel networks $G_1$ and $G_2$ in parallel.\\
Same as case 1, we denote $\Rp$ (resp. $\Rpp$) as the set of routes in $\Gp$ (resp. $\Gpp$). Then, the set of all routes in $\G$ is $\R = \Rp \cup \Rpp$. 

Given any $\fhat$, we compute $\xopt$ from the procedure (i) -- (iii) in network $\G$. We denote $\foptp = \sum_{\rp \in \Rp} \sum_{\b \in \B} \xrbopt$ (resp. $\foptpp = \sum_{\rpp \in \Rpp} \sum_{\b \in \B} \xrbopt$) as the total flow assigned to subnetwork $\Gp$ (resp. $\Gpp$) given $\xopt$. We now denote $\xoptp$ (resp. $\xoptpp$) as the trip vector $\xopt$ restricted on sub-network $\Gp$ (resp. $\Gpp$), i.e. $\xoptp = \(\xopt_{\rp}(\b)\)_{\rp \in \Rp, \b \in \B}$ (resp. $\xoptpp = \(\xopt_{\rpp}(\b)\)_{\rpp \in \Rpp, \b \in \B}$). We can check that $\xoptp$ (resp. $\xoptpp$) is the trip  vector obtained by the procedure (i) -- (iii) given the total flow $\foptp$ (resp. $\foptpp$) on network $\Gp$ (resp. $\Gpp$). 

Consider any arbitrary split of the total flow $\fhat$ to the two sub-networks, denoted as $\(\fhatp, \fhatpp\)$, such that $\fhatp(\b)+\fhatpp(\b)=\fhat(\b)$ for all $\b \in \B$. Given $\fhatp$ (resp. $\fhatpp$), we denote the trip  vector obtained by procedure (i) -- (iii) on sub-network $\Gp$ (resp. $\Gpp$) as $\xopthatp$ (resp. $\xopthatpp$). We also define the set of feasible trip  vectors on sub-network $\Gp$ (resp. $\Gpp$) that induce the total flow $\fhatp$ (resp. $\fhatpp$) given by \eqref{eq:X} as $\Xp(\fhatp)$ (resp. $\Xpp(\fhatpp)$). Then, the set of all trip  vectors that induce $\fhat$ on network $G$ is $\X(\fhat) = \cup_{\(\fhatp, \fhatpp\)} (\Xp(\fhatp), \Xpp(\fhatpp))$. 

Under our assumption that \eqref{eq:induction} holds on sub-network $\Gp$ and $\Gpp$ with any total flow, we know that given any flow split $\(\fhatp, \fhatpp\)$, 
\begin{align*}
    \sum_{\rp \in \Rp}  \sum_{\b \in \B} g(\b)\tr \xopthatp_{\rp}(\b) + \sum_{\rpp \in \Rpp}  \sum_{\b \in \B} g(\b)\tr \xopthatpp_{\rpp}(\b) \leq \sum_{\rp \in \Rp}  \sum_{\b \in \B} g(\b)\tr \xhatp_{\rp}(\b) &+\sum_{\rpp \in \Rpp}  \sum_{\b \in \B} g(\b)\tr \xhatpp_{\rpp}(\b), \\
    &\quad \forall \xhatp \in \X(\fhatp), \xhatpp \in \X(\fhatpp).
\end{align*}
Therefore, the optimal solution of \eqref{eq:induction} must be a trip  vector $\(\xopthatp, \xopthatpp\)$ associated with a flow split $\(\fhatp, \fhatpp\)$. It thus remains prove that any $\(\xopthatp, \xopthatpp\)$ associated with flow split $\(\fhatp, \fhatpp\) \neq \(\foptp, \foptpp\)$ cannot be an optimal solution (i.e. can be improved by re-arranging flows). 

For any $\(\fhatp, \fhatpp\) \neq \(\foptp, \foptpp\)$, we can find a group $\bj$ such that $\fhatp(\bj) \neq \foptp(\bj)$ (henceforth $\fhatpp(\bj) \neq \foptpp(\bj)$). We denote $\bjhat$ as one such group with the maximum $g(\b)$, i.e. $\fhatp(\bj)= \foptp(\bj)$ for any $j -1, \dots, \jhat-1$. Since groups $\b_1, \dots, \b_{\jhat-1}$ are assigned before group $\bjhat$ according to procedure (i) -- (iii), we know that $\xoptp_{\rp}(\bj)=\xopt_{\rp}(\bj)$ and $\xoptpp_{\rpp}(\bj)=\xopt_{\rpp}(\bj)$ for all $\rp \in \Rp$, all $\rpp \in \Rpp$ and all $j=1, \dots, \jhat-1$. Since $\fhatp(\bjhat) \neq \foptp(\bjhat)$, the trip  vector in $\xoptp$ and $\xoptpp$ must be different from that in $\xopt$. Without loss of generality, we assume that $\fhatp(\bjhat)> \foptp(\bjhat)$ and $\fhatpp(\bjhat)< \foptpp(\bjhat)$. Then, there must exist routes $\rphat \in \Rp$ and $\rpphat \in \Rpp$ such that $\xoptp_{\rphat}(\bjhat) > \xopt_{\rphat}(\bjhat)$ and $\xoptpp_{\rpphat}(\bjhat) < \xopt_{\rpphat}(\bjhat)$. Moreover, since $\xopt$ assigns group $\bjhat$ to routes with the minimum travel time cost that are unsaturated after assigning groups $b_1, \dots, \b_{\jhat-1}$, we have $t_{\rpphat} < t_{\rphat}$. If route $\rpphat$ is unsaturated given $\xopthatpp$, then we decrease $\xopthatp_{\rphat}(\bjhat)$ and increase $\xopthatpp_{\rpphat}(\bjhat)$ by a small positive number $\epsilon>0$. We can check that the objective function of \eqref{eq:induction} is reduced by $\epsilon (t_{\rphat}- t_{\rpphat})\epsilon g(\bjhat) > 0$. On the other hand, if route $\rpphat$ is saturated, then group $\b_{\jhat+1}$ must be assigned to $\rpphat$ because it is assigned right after group $\bjhat$. Then, we decrease $\xoptp_{\rphat}(\bjhat)$ and $\xoptpp_{\rpphat}(\b_{\jhat+1})$ by $\epsilon>0$, increases $\xoptp_{\rphat}(\b_{\jhat+1})$ and $\xoptpp_{\rpphat}(\b_{\jhat})$ by $\epsilon$ (i.e. exchange a small fraction of group $\bjhat$ with group $\b_{\jhat+1}$). Note that $g(\bjhat)> g(\b_{\jhat+1})$ and $t_{\rphat}> t_{\rpphat}$. We can thus check that the objective function of \eqref{eq:induction} is reduced by $\epsilon (t_{\rphat} g(\bjhat)- t_{\rpphat}g(\b_{\jhat+1}))\epsilon > 0$. Therefore, we have found an adjustment of trip  vector $\(\xopthatp, \xopthatpp\)$ that reduces the objective function of \eqref{eq:induction}. Hence, for any flow split $\(\fhatp, \fhatpp\) \neq \(\foptp, \foptpp\)$, the associated trip  vector $\(\xopthatp, \xopthatpp\)$ is not the optimal solution of \eqref{eq:induction}. The optimal solution of \eqref{eq:induction} must be constructed by procedure (i) -- (iii) with flow split $\(\foptp, \foptpp\)$, i.e. must be $\xopt$. 

We have shown from cases 1 and 2 that if $\xopt$ is an optimal solution of \eqref{eq:induction} on two series-parallel sub-networks, then $\xopt$ is an optimal solution on the connected series-parallel network. Moreover, since \eqref{eq:induction} holds trivially when the network is a single edge, and any series-parallel network is formed by connecting series-parallel sub-networks in series or parallel, we can conclude that $\xopt$ obtained from procedure (i) -- (iii) minimizes the objective function in \eqref{eq:induction} for any flow vector $\fhat$ on any series-parallel network.  

From \eqref{eq:decompose_V}, \eqref{eq:equal_part} and \eqref{eq:induction}, we can conclude that $V(\xopt) \geq V(\xhatopt)$. Hence, $\xopt$ must be an optimal solution in \eqref{eq:LP2k}.  
\QEDA

\vspace{0.2cm}
\noindent\emph{Proof of Lemma \ref{lemma:x_convert_y}.} First, for any feasible $\x$ in \eqref{eq:LP2k}, consider a vector $\yall$ such that for any $\(\r, \b\) \in \{\B \times \R|\xrb=1\}$, $\yall_l(\b)=1$ for one $\l \in \Lr$ and $\yall_l(\ball)=0$ for any other $\(\ball, \l\)$. We can check that $\yall$ is feasible in \eqref{eq:opt_lane} and $S(\x)=S(\yall)$. On the other hand, for any feasible $\y$ in \eqref{eq:opt_lane}, there exists $\x=\chi(\y)$ as in \eqref{eq:chi} such that $\x$ is feasible in \eqref{eq:LP2k} and $S(\x)=S(\yall)$. Thus, \eqref{eq:LP2k} and \eqref{eq:opt_lane} are equivalent in that for any feasible solution of one linear program, there exists a feasible solution that achieves the same social welfare in the other linear program. 

Therefore, \eqref{eq:LP2k} has an integer optimal solution if and only if \eqref{eq:opt_lane} has an integer optimal solution, and for any integer optimal solution $\yopt$ of \eqref{eq:opt_lane}, $x=\chi(\yopt)$ as in \eqref{eq:chi} is an optimal solution of \eqref{eq:LP2k}. \QEDA

\vspace{0.3cm}

\noindent\emph{Proof of Lemma \ref{lemma:equivalent_economy}.} We write the dual program of \eqref{eq:opt_lane} as follows: 
\begin{subequations}
        \makeatletter
        \def\@currentlabel{D-y}
        \makeatother
        \label{eq:Dlane}
        \renewcommand{\theequation}{D-y.\alph{equation}}
    \begin{align}
         \min_{\u, \mu}  \quad & \sum_{\m \in \M} \um + \sum_{\l \in \L} \mu_l , \notag\\
   s.t. \quad  &  \sum_{\m \in \ball} \um + \mu_l \geq W_l(\ball) \quad \forall \ball \in \Ball, \quad \forall \l \in \L,  \label{subeq:Dlane1}\\
    &\um \geq 0, \mu_l \geq 0, \quad \forall \m \in \M, \quad \forall \l \in \L.\label{subeq:Dlane2}
    \end{align}
\end{subequations}
For any Walrasian equilibrium $\(\yopt, \uopt\)$, we consider the vector $\muopt=\(\muopt_l\)_{\l \in \L}$ as follows: 
\begin{align}\label{eq:muopt}
    \muopt_l = \max_{\ball \in \Ball} W_l(\ball) - \sum_{\m \in \ball} \umopt, \quad \forall \l \in \L. 
\end{align}
From the definition of Walrasian equilibrium, we know that $\yopt$ is a feasible solution of \eqref{eq:opt_lane}, and $\(\uopt, \muopt\)$ is a feasible solution of \eqref{eq:Dlane}. We now show that $\(\yopt, \uopt, \muopt\)$ satisfies complementary slackness condition of \eqref{eq:opt_lane} and \eqref{eq:Dlane}. 
\begin{itemize}
    \item[-] Complementary slackness condition for \eqref{eq:lane_1}: Condition \emph{(ii)} in Definition \ref{def:gross_substitute} ensures that rider $\m$'s utility is positive if and only if \eqref{eq:lane_1} is tight (i.e. rider $\m$ joins a trip). 
    \item[-] Complementary slackness condition for \eqref{eq:lane_2}: If no rider group takes route $\l \in \L$, i.e. \eqref{eq:lane_2} is slack and $\bl=0$, then $\muopt_l$ as in \eqref{eq:muopt} is zero. On the other hand, $\muopt_l>0$, then $\bl \neq 0$. Hence, \eqref{eq:lane_2} must be tight.
    \item[-] Complementary slackness condition for \eqref{subeq:Dlane1}: From condition (i) in Definition \ref{def:gross_substitute}, we know that $\yopt_l(\bl)=1$ if and only if $\bl \in \argmax_{\ball \in \Ball} W_l(\ball) - \sum_{\m \in \ball} \umopt$, i.e. constraint \eqref{subeq:Dlane1} is tight.  
\end{itemize}
From strong duality, we know that $y^*$ must be an integer optimal solution of \eqref{eq:opt_lane} and $\(\u^*, \muopt\)$ must be an optimal solution of \eqref{eq:Dlane}. Therefore, we can conclude that a Walrasian equilibrium $\(y^*t, \uopt\)$ exists in the equivalent economy $\econ$ if and only if \eqref{eq:opt_lane} has an optimal integer solution. \QEDA

\vspace{0.3cm}
\noindent\emph{Proof of Lemma \ref{lemma:condition_gross}.} Since all riders have homogeneous carpool disutility, we can simplify the trip value function from \eqref{eq:value_of_trip_tilde} as follows: 
\begin{align*}
    \Wrb = \sum_{\m \in \tildeb} \etamr -\thetafun(|\tildeb|) \tr, 
\end{align*}
where $\etamr \deleq \tripm - \vm\tr$ and $\thetafun(|\tildeb|)= |\tildeb|\gamma(|\tildeb|) + \unitcost(|\tildeb|)$. 

Before proving that the augmented trip value function $\W_r(\ball)$ satisfies (a) and (b) in Definition \ref{def:gross_substitute}, we first prove the following statements that will be used later: 

\emph{(i)} The function $\thetafun(|\rep(\ball)|)$ is non-decreasing in $|\rep(\ball)|$ because the marginal carpool disutility is non-decreasing in the group size. 

\emph{(ii)} The representative rider group for any trip $\(\ball, \r\) \in \Ball \times \R$ can be constructed by selecting riders from $\ball$ in decreasing order of $\etamr$. The last selected rider $\ell$ (i.e. the rider in $\rep(\ball)$ with the minimum value of $\etamr$) satisfies: 
\begin{align}\label{eq:ell_one}
    \eta_r^{\ell} \geq \(\thetafun(|\rep(\ball)|) - \thetafun(|\rep(\ball)|-1)\)\tr.
\end{align}
That is, adding rider $\ell$ to the set $\rep(\ball) \setminus \{\ell\}$ increases the trip valuation. Additionally, 
\begin{align}\label{eq:ell_two}
    \etamr < \(\thetafun(|\rep(\ball)|+1) - \thetafun(|\rep(\ball)|)\)\tr, \quad \forall \m \in \ball \setminus \rep(\ball).
\end{align}
Then, adding any rider in $\ball \setminus \rep(\ball)$ to $\rep(\ball)$ no longer increases the trip valuation.

\emph{(iii)} $|\rep(\ball')| \geq |\rep(\ball)|$ for any two rider groups $\ball', \ball \in \B$ such that $\ball' \supseteq \ball$. \\
\emph{Proof of (iii).} Assume for the sake of contradiction that $|\rep(\ball')|< |\rep(\ball)|$. Consider the rider $\ell \in \argmin_{\m \in \rep(\ball)} \eta_r^m$. The value $\eta_r^{\ell}$ satisfies  \eqref{eq:ell_one}. Since $|\rep(\ball')|< |\rep(\ball)|$, $\ball' \supseteq \ball$, and we know that riders in the representative rider group $\rep(\ball')$ are the ones with $|\rep(\ball')|$ highest $\eta_r^m$ in $\ball'$, we must have $\ell \notin \rep(\ball')$. From \eqref{eq:ell_two}, we know that $\eta_r^{\ell} < \(\thetafun(|\rep(\ball')|+1) - \thetafun(|\rep(\ball')|)\)\tr$. Since the marginal carpool disutility is non-decreasing in the rider group size, we can check that $\thetafun(|\rep(\ball)|+1) - \thetafun(|\rep(\ball)|)$ is non-decreasing in $|\rep(\ball)|$. Since $|\rep(\ball')|< |\rep(\ball)|$, we have $|\rep(\ball')|\leq |\rep(\ball)|-1$. Therefore, \[\eta_r^{\ell} < \(\thetafun(|\rep(\ball')|+1) - \thetafun(|\rep(\ball')|)\)\tr \leq \(\thetafun(|\rep(\ball)|) - \thetafun(|\rep(\ball)|-1)\)\tr,\] which contradicts \eqref{eq:ell_one} and the fact that $\ell \in \tildeb$. Hence, $|\rep(\ball')| \geq |\rep(\ball)|$. 

We now prove that $\W$ satisfies \emph{(i)} in Definition \ref{def:gross_substitute}.
For any $\ball, \ball' \subseteq \M$ and $\ball \subseteq \ball'$, consider two cases: \\
\emph{Case 1:} $i \notin \rep(\{i\} \cup\ball')$. In this case, $\rep(\ball' \cup i) = \rep(\ball')$, and $\W(i|\ball')=\W(\ball' \cup i) - \W(\ball')=0$. Since $\W$ satisfies monotonicity condition, we have $\W(i|\ball) \geq 0 $. Therefore, $\W(i|\ball) \geq \W(i|\ball')$.

\noindent\emph{Case 2:} $i \in \rep(\{i\} \cup\ball')$. We argue that $i \in \rep(\{i\} \cup\ball)$. From 
\eqref{eq:ell_one}, $\eta_r^i \geq \(\thetafun(|\rep(\ball')|) - \thetafun(|\rep(\ball')|-1)\)\tr$. Since $\ball' \supseteq \ball$, we know from (iii) that $|\rep(\ball')| \geq |\rep(\ball)|$. Hence, $\eta_r^i \geq \(\thetafun(|\rep(\ball)|) - \thetafun(|\rep(\ball)|-1)\)\tr$, and thus $i \in \rep(\{i\} \cup\ball)$. 

We define $\ell' \deleq \argmin_{m \in \rep(\ball')} \eta_r^m$ and $\ell \deleq \argmin_{m \in \rep(\ball)} \eta_r^m$. We also consider two thresholds $\mu' = \(\thetafun(|\rep(\ball')|+1) - \thetafun(|\rep(\ball')|)\)\tr$, and $\mu = \(\thetafun(|\rep(\ball)|+1) - \thetafun(|\rep(\ball)|)\)\tr$.  
Since $\ball' \supseteq \ball$, from (iii), we have $|\rep(\ball')| \geq |\rep(\ball)|$ and thus $\mu' \geq \mu$. 
We further consider four sub-cases:

\emph{(2-1)} $\eta_r^{\ell'} \geq \mu'$ and $\eta_r^{\ell} \geq \mu$. From \eqref{eq:ell_one} and \eqref{eq:ell_two}, $\rep(\{i\} \cup\ball') = \rep(\ball') \cup \{i\}$ and $\rep(\{i\} \cup\ball) = \rep(\ball) \cup \{i\}$. The marginal value of $i$ is $\W_r(i |\ball')=\eta_r^i - \mu'$, and $\W_r(i|\ball)= \eta_r^i- \mu$. Since $\mu' \geq \mu$, $\W_r(i|\ball') \leq \W_r(i|\ball)$. 
    
\emph{(2-2)} $\eta_r^{\ell'}< \mu'$ and $\eta_r^{\ell} \geq \mu$. Since $i \in \rep(\{i\} \cup\ball')$ in \emph{Case 2}, we know from \eqref{eq:ell_one} and \eqref{eq:ell_two} that $\rep(\{i\} \cup\ball')= \rep(\ball') \setminus \{\ell'\}\cup\{i\}$ and $\rep(\{i\} \cup\ball) = \rep(\ball) \cup \{i\}$. Therefore, $\W_r(i|\ball') = \eta_r^i -\eta_r^{\ell'}$ and $\W_r(i|\ball) = \eta_r^i- \mu$. We argue in this case, we must have $|\rep(\ball')|>|\rep(\ball)|$. Assume for the sake of contradiction that $|\rep(\ball')|=|\rep(\ball)|$, then $\mu' = \mu$ and $\eta_r^{\ell'} \geq \eta_r^{\ell}$ because $\ball' \supseteq \ball$. However, this contradicts the assumption of this subcase that $\eta_r^{\ell'} < \mu' = \mu \leq \eta_{r}^{\ell}$. Hence, we must have $|\rep(\ball')| \geq |\rep(\ball)|+1$. Then, from \eqref{eq:ell_one}, we have $\eta_r^{\ell'} \geq \(\thetafun(|\rep(\ball')|) - \thetafun(|\rep(\ball')|-1)\)\tr \geq \mu$. Hence, $\W_r(i|\ball')  \leq \W_r(i|\ball)$. 

\emph{(2-3)} $\eta_r^{\ell'}\geq \mu'$ and $\eta_r^{\ell} < \mu$. From \eqref{eq:ell_one} and \eqref{eq:ell_two}, $\rep(i \cup\ball') = \rep(\ball') \cup \{i\}$ and $\rep(\{i\} \cup\ball) = \rep(\ball) \setminus \{\ell'\}  \cup \{i\}$. Therefore, $\W_r(i|\ball') = \eta_r^i -\mu'$ and $\W_r(i|\ball) = \eta_r^i- \eta_r^{\ell}$. Since $\mu' \geq \mu \geq \eta_r^{\ell}$, we know that $\W_r(i|\ball')  \leq \W_r(i|\ball)$. 

\emph{(2-4)} $\eta_r^{\ell'} < \mu'$ and $\eta_r^{\ell} < \mu$. From \eqref{eq:ell_one} and \eqref{eq:ell_two},  $\rep(\{i\} \cup\ball') = \rep(\ball') \setminus \{\ell'\}\cup \{i\}$, and $\rep(\{i\} \cup\ball) = \rep(\ball) \setminus \{\ell\}\cup \{i\}$. Therefore, $\W_r(i|\ball') = \eta_r^i -\eta_r^{\ell'}$ and $\W_r(i|\ball) = \eta_r^i- \eta_r^{\ell}$. If $|\rep(\ball')| = |\rep(\ball)|$, then we must have $\eta_r^{\ell'} \geq \eta_r^{\ell}$, and hence $\W_r(i|\ball') \leq \W_r(i|\ball)$. On the other hand, if $|\rep(\ball')| \geq |\rep(\ball)|+1$, then from \eqref{eq:ell_one} we have $\eta_r^{\ell} \geq \(\thetafun(|\rep(\ball')|) - \thetafun(|\rep(\ball')|-1)\)\tr \geq \mu > \eta_r^{\ell}$. Therefore, we can also conclude that $\W_r(i|\ball') \leq \W_r(i|\ball)$.

From all four subcases, we can conclude that in case 2, $\W_r(i|\ball) \geq \W_r(i|\ball')$.  

We now prove that $\W$ satisfies condition (\emph{ii}) of Definition \ref{def:gross_substitute} by contradiction. Assume for the sake of contradiction that \eqref{eq:gs} is not satisfied. Then, there must exist a group $\ball \in \Ball$, and $i, j,k \in M\setminus \ball$ such that: 
\begin{subequations}
\begin{align}
    &\W_r(i, j|\ball)+ \W_r(k|\ball) > \W_r(i|\ball) + \W_r(j, k|\ball),\quad \Rightarrow \quad \W_r(j|i, \ball) > \W_r(j|k, \ball), \label{subeq:one}\\
    &\W_r(i, j|\ball)+ \W_r(k|\ball) > \W_r(j|\ball) + \W_r(i,k|\ball), \quad \Rightarrow \quad \W_r(i|j, \ball) > \W_r(i|k, \ball).\label{subeq:two}
\end{align}
\end{subequations}
We consider the following four cases: 

\emph{Case A:} $\rep\(\ball \cup \{i, j\}\) = \rep\(\ball \cup \{i\}\) \cup \{j\}$ and $\rep\(\ball \cup \{j, k\}\) = \rep\(\ball \cup \{k\}\) \cup \{j\}$. In this case, if $|\rep\(\ball \cup \{i\}\)| \geq  |\rep\(\ball \cup \{k\}\)|$, then $\W_r(j|i, \ball) \leq \W_r(j|k, \ball)$, which contradicts \eqref{subeq:one}. On the other hand, if $|\rep\(\ball \cup \{i\}\)| <  |\rep\(\ball \cup \{k\}\)|$, then we must have $\rep\(\ball \cup \{i\}\)=\rep(\ball)$ and $\rep\(\ball \cup \{k\}\) = \rep(\ball) \cup \{k\}$. Therefore, $\W_r(i|j, \ball)=0$, and \eqref{subeq:two} cannot hold. We thus obtain the contradiction.  

\emph{Case B:} $|\rep\(\ball \cup \{i, j\}\)| = |\rep\(\ball \cup \{i\}\)|$ and $|\rep\(\ball \cup \{j, k\}\) = \rep\(\ball \cup \{k\}\)|$. We further consider the following four sub-cases: 

\emph{(B-1).} $\rep\(\ball \cup \{i, j\}\) = \rep\(\ball \cup \{i\}\) $ and $\rep\(\ball \cup \{j, k\}\) = \rep\(\ball \cup \{k\}\) $. In this case, $\W_r(j|i, \ball) = \W_r(j|k, \ball)=0$. Hence, we arrive at a contradiction against \eqref{subeq:one}. 

\emph{(B-2).} $\rep\(\ball \cup \{i, j\}\) \neq \rep\(\ball \cup \{i\}\) $ and $\rep\(\ball \cup \{j, k\}\) = \rep\(\ball \cup \{k\}\) $. In this case, when $j$ is added to the set $\ball \cup \{i\}$, $j$ replaces a rider, denoted as $\ell \in \ball \cup \{i\}$. Since $\ell$ is replaced, we must have $\eta^{\ell}_r \leq \eta^m_r$ for any $\m \in \rep(\ball \cup\{j\})$. If $\ell = i$, then $\rep(\ball \cup\{i, j\}) = \rep(\ball \cup \{j\})$. Hence, $\W_r(i|j, \ball)=0$, and we arrive at a contradiction with \eqref{subeq:two}. On the other hand, if $\ell \neq i$, then $\ell$ is a rider in group $\ball$. This implies that $\ell \in \ball$ should be replaced by $j$ when $j$ is added to the set $\{k\} \cup \ball$, which contradicts the assumption of this case that $\rep\(\ball \cup \{j, k\}\) = \rep\(\ball \cup \{k\}\) $. 
        
\emph{(B-3).} $\rep\(\ball \cup \{i, j\}\) = \rep\(\ball \cup \{i\}\) $ and $\rep\(\ball \cup \{j, k\}\) \neq \rep\(\ball \cup \{k\}\)$. Analogous to case \emph{B-2}, we know that $\rep\(\ball \cup \{j, k\}\) = \rep\(\ball \cup \{j\}\)$ and $\eta_r^j \geq \eta^k_r$.  Moreover, since $\rep\(\ball \cup \{i, j\}\) = \rep\(\ball \cup \{i\}\) $, we must have $\eta_r^j \leq \eta^i_r$. Therefore, $\W_r(\ball \cup \{i, j\}) = \W_r(\ball \cup \{i\})$, and $\W_r(i|j, \ball)= \W_r(\ball\cup \{i\}) - \W_r(\ball\cup \{j\})$. Since $\eta_r^j \leq \eta^i_r$ and $\eta_r^j \geq \eta^k_r$, we know that $\W_r(i|k, \ball) =\W_r(\ball\cup \{i\}) - \W_r(\ball \cup \{k\}) \geq \W_r(\ball\cup \{i\}) - \W_r(\ball\cup \{j\}) = \W_r(i|j, \ball)$, which contradicts \eqref{subeq:two}. 

\emph{(B-4).} $\rep\(\ball \cup \{i, j\}\) \neq \rep\(\ball \cup \{i\}\) $ and $\rep\(\ball \cup \{j, k\}\) \neq \rep\(\ball \cup \{k\}\) $. In this case, if $\rep\(\ball \cup \{i, j\}\) = \rep\(\ball \cup \{j\}\)$, then $\W_r(i|j, \ball)= \W_r(i, j, \ball)- \W_r(j, \ball) = \W_r(j, \ball) - \W_r(j, \ball)=0$, which contradicts \eqref{subeq:two}. On the other hand, if $\rep\(\ball \cup \{i, j\}\) \neq \rep\(\ball \cup \{j\}\)$, then one rider $\ell \in \ball$ must be replaced by $j$ when $j$ is added into the set $\ball \cup \{i\}$, i.e. $\rep\(\ball \cup \{i, j\}\) = \rep\(\ball\setminus \{\ell\} \cup \{i, j\}\) $. Hence, $\eta_r^{\ell} \leq \eta_r^i$ and $\eta_r^{\ell} \leq \eta_r^{j}$. If $\eta_r^{\ell} \leq  \eta_r^k$, then under the assumption that $|\rep\(\ball \cup \{j, k\}\)|=|\rep\(\ball \cup \{k\}\)|$ and $\rep\(\ball \cup \{j, k\}\) \neq \rep\(\ball \cup \{k\}\)$, we must have $\rep\(\ball \cup \{j, k\}\)= \rep\(\ball\setminus \{\ell\} \cup \{j, k\}\)$. Then, we can check that $\W_r(j|i, b) = \W_r(j|k, b)$, which contradicts \eqref{subeq:one}. 

        On the other hand, if $\eta_r^{\ell} > \eta_r^k$, then $\rep\(\ball \cup \{j, k\}\)= \rep\(\ball \cup \{j\}\)$. In this case, $\W_r(i|j, \ball)$ is the change of trip value by replacing $\ell$ with $i$, and $\W_r(i|k, \ball)$ is the change of trip value by replacing $k$ with $i$. Since $\eta_r^k < \eta_r^{\ell}$, we must have $\W_r(i|j, \ball) < \W_r(i|k, \ball)$, which contradicts \eqref{subeq:two}.

\emph{Case C:} $\rep\(\ball \cup \{i, j\}\) = \rep\(\ball \cup \{i\}\) \cup \{j\}$ and $|\rep\(\ball \cup \{j, k\}\)| = |\rep\(\ball \cup \{k\}\)|$. We further consider the following sub-cases: 

\emph{(C-1).}  $\rep\(\ball \cup \{j, k\}\) = \rep\(\ball \cup \{k\}\)$. In this case, $\eta_r^j \leq \eta_r^m$ for all $m \in \rep(\ball \cup \{k\})$, and $\eta_r^j < \thetafun(|\rep(\ball \cup \{k\})+1|) - \thetafun(|\rep(\ball \cup \{k\})|)$. Since $\rep\(\ball \cup \{i, j\}\) = \rep\(\ball \cup \{i\}\) \cup \{j\}$, we know that $\eta_r^j \geq \thetafun(|\rep(\ball \cup \{i\})+1|) - \thetafun(|\rep(\ball \cup \{i\})|)$. Since carpool disutility is non-decreasing in rider group size, for $\eta_r^j$ to satisfy both inequalities, we must have $|\rep(\ball \cup \{i\})| < |\rep(\ball \cup \{k\})|$. Then, we must have $\rep(\ball\cup \{i\}) = \rep(\ball)$ and $\rep(\ball\cup \{k\}) = \rep(\ball) \cup \{k\}$. Therefore, $\W_r(i, j, \ball) = \W_r(j, \ball)$ and $\W_r(i, k, \ball) = \W_r(k, \ball)$. Hence, $\W_r(i|j, \ball) = \W_r(i|k, \ball)=0$, which contradicts \eqref{subeq:two}. 

\emph{(C-2).}  $\rep\(\ball \cup \{j, k\}\) \neq \rep\(\ball \cup \{k\}\)$. Since $|\rep\(\ball \cup \{j, k\}\)| = |\rep\(\ball \cup \{k\}\)|$, $j$ replaces a rider $\ell$ in $\ball \cup \{k\}$, and $\eta^{\ell}_{\r} \leq \ell_r^{\m}$ for all $\m \in \ball \cup{k}$. If $\ell = k$, then $\rep\(\ball \cup \{j, k\}\) = \rep\(\ball \cup \{j\}\)$. Therefore, $\W_r(j|i, \ball) = \eta_r^{j} - \(\thetafun(|\rep(\ball \cup \{i\})|+1) - \thetafun(|\rep(\ball \cup \{i\})|)\)$ and $\W_r(j|k, \ball) = \eta_r^{j} - \eta_r^{k}$. If $\eta_r^k \leq \thetafun(|\rep(\ball \cup \{i\})|+1) - \thetafun(|\rep(\ball \cup \{i\})|)$, then \eqref{subeq:one} is contradicted. Thus, $\eta_r^k > \thetafun(|\rep(\ball \cup \{i\})|+1) - \thetafun(|\rep(\ball \cup \{i\})|)$. Since $k$ is replaced by $j$ when $j$ is added to $\ball \cup \{k\}$, we must have $\eta_r^k < \thetafun(|\rep(\ball \cup \{j\})|+1) - \thetafun(|\rep(\ball \cup \{j\})|)$. For $\eta_r^k$ to satisfy both inequalities, we must have $|\rep(\ball \cup \{j\})| > |\rep(\ball \cup \{i\})|$. Hence, $\rep(\ball \cup \{j\}) = \rep(\ball) \cup \{j\}$ and $\rep(\ball \cup \{i\}) = \rep(\ball)$. Then, $\W_r(i|j, \ball) = \W_r(\ball\cup \{i, j\}) -  \W_r(\ball\cup \{ j\}) = 0$, which contradicts \eqref{subeq:two}. 

On the other hand, if $\ell \in \ball$, then we know from \eqref{eq:ell_two} that $\eta_r^{\ell} < \thetafun(|\rep\(\ball \cup \{k\}\)|+1) - \thetafun(|\rep\(\ball \cup \{k\}\)|)$. Additionally, since $\rep\(\ball \cup \{i, j\}\) = \rep\(\ball \cup \{i\}\) \cup \{j\}$, we know from \eqref{eq:ell_one} that $\eta_r^{\ell} \geq \thetafun(|\rep\(\ball \cup \{i\}\)|+1) - \thetafun(|\rep\(\ball \cup \{i\}\)|)$. If $\eta_r^{\ell}$ satisfies both inequalities, then we must have $|\rep\(\ball \cup \{i\}\)|< |\rep\(\ball \cup \{k\}\)|$. Therefore, $\rep\(\ball \cup \{i\}\)= \rep(\ball)$. Then, $\W_r(i|j, \ball)=0$, which contradicts \eqref{subeq:two}.

\emph{Case D:} $|\rep\(\ball \cup \{i, j\}\)| = |\rep\(\ball \cup \{i\}\)|$ and $\rep\(\ball \cup \{j, k\}\) = \rep\(\ball \cup \{k\}\) \cup \{j\}$. We further consider the following sub-cases: 

\emph{(D-1).}  $\rep\(\ball \cup \{i, j\}\) = \rep\(\ball \cup \{i\}\)$. In this case, analogous to \emph{(C-1)}, we know that $|\rep(\ball \cup \{k\})| < |\rep(\ball \cup \{i\})|$. Therefore, $\rep(\ball\cup \{k\})=\rep(\ball)$ and $\rep(\ball \cup \{i\})=\rep(\ball) \cup \{i\}$. Therefore, $\eta_r^k < \eta_r^i$. Additionally, since $\rep\(\ball \cup \{i, j\}\) = \rep\(\ball \cup \{i\}\)$, $\eta^j_r < \eta_r^i$. Then, $\W_r(i|j, \ball)= \W_r(i, \ball) - \W_r(j, \ball)$ and $\W_r(i|k, \ball) = \W_r(i, \ball)- \W_r(\ball)$. Since $\W$ is monotonic, $\W_r(j, \ball) \geq \W_r(\ball)$ so that $\W_r(i|j, \ball) \leq \W_r(i|k, \ball)$, which contradicts \eqref{subeq:two}. 

\emph{(D-2).} $\rep\(\ball \cup \{i, j\}\) \neq \rep\(\ball \cup \{i\}\)$. Since $|\rep\(\ball \cup \{i, j\}\)| = |\rep\(\ball \cup \{i\}\)|$, $j$ replaces the rider $\ell \in \ball \cup \{i\}$ such that $\eta^{\ell}_r \leq \eta^{\m}_r$ for all $\m \in \rep(\ball \cup \{i\})$. If $\ell = i$, then analogous to case \emph{C-2}, we know that if \eqref{subeq:two} is satisfied, then $|\rep(\ball \cup \{j\})|< |\rep(\ball \cup \{k\})|$. Hence, $\rep(\ball \cup \{j\}) = \rep(\ball)$ and $V(j |i, \ball) =0$, which contradicts \eqref{subeq:one}. 

On the other hand, if $\ell \in \ball$, then again analogous to case \emph{C-2}, we know that $|\rep\(\ball \cup \{k\}\)|< |\rep\(\ball \cup \{i\}\)|$. Therefore, $\rep\(\ball \cup \{k\}\)=\rep(\ball)$, and $\rep\(\ball \cup \{i\}\)= \rep(\ball)\cup \{i\}$. Then, $\W_r(j|i, \ball) = \W_r(\ball \setminus \{\ell\} \cup \{i, j\}) - \W_r(i, \ball)$, and $\W_r(j|k, \ball) = \W_r(\ball\cup \{j\}) - \W_r(\ball)$. Since $\ell \neq i$, $\W_r(i|j, \ball)=\W_r(\ball \setminus \{\ell\} \cup \{i, j\}) - \W_r(j, \ball) = \eta_r^{i}- \eta_r^{\ell}$. Additionally, since $\rep(i, \ball) = \rep(\ball) \cup \{i\}$, $\W_r(i|k, \ball)=\W_r(i, \ball) - \W_r(\ball) = \eta_r^{i} -(\thetafun(|\rep(\ball)|+1)- \thetafun(|\rep(\ball)|))$. Since $\rep(\ball \cup \{i\})= \rep(\ball) \cup \{i\}$ and $\ell \in \ball$, we know from \eqref{eq:ell_one} that $\eta_r^{\ell}\geq  \thetafun(|\rep(\ball)|+1)- \thetafun(|\rep(\ball)|)$. Therefore, $\W_r(i|j, \ball) \leq \W_r(i|k, \ball)$, which contradicts \eqref{subeq:two}.

From all above four cases, we can conclude that condition \emph{(ii)} of Definition \ref{def:gross_substitute} is satisfied. We can thus conclude that $\W$ satisfies gross substitutes condition.
\QEDA

\section{Proof of Section \ref{sec:strategyproof}}\label{apx:proof_C}
\noindent\emph{Proof of Lemma \ref{lemma:utility}.} We first show that for any optimal utility vector $\uopt \in \Uopt$, there exists a vector $\lambda^*$ such that $\(\uopt, \lambda^*\)$ is an optimal solution of \eqref{eq:D2k}. Since $\uopt \in \Uopt$, there must exist a toll price vector $\tollopt$ such that $\(\uopt, \tollopt\)$ is an optimal solution of \eqref{eq:D1bar}. Consider $\lambda^* = \(\lambda^*_r\)_{\r \in \Ropt}$ as follows: 
\begin{align}\label{eq:lambdaopt}
\lambda^*_r = \sum_{\e \in \r} \tollopt_e, \quad \forall \r \in \Ropt. 
\end{align}
Since $\(\uopt, \tollopt\)$ is feasible in \eqref{eq:D1bar}, we can check that $\(\uopt, \lambda^*\)$ is also a feasible solution of \eqref{eq:D2k}. Moreover, since $\(\xopt, \uopt, \tollopt\)$ satisfies complementary slackness conditions with respect to \eqref{eq:LP1bar} and \eqref{eq:D1bar}, $\(\xopt, \uopt, \lambda^*\)$ also satisfies complementary slackness conditions with respect to \eqref{eq:LP2k} and \eqref{eq:D2k}. Therefore, $\(\umopt, \lambda^*\)$ is an optimal solution of \eqref{eq:D2k}. 

We next show that for any optimal solution $\(\uopt, \lambda^*\)$ of \eqref{eq:D2k}, we can find a toll price vector $\tollopt$ such that $\(\uopt, \tollopt\)$ is an optimal solution of \eqref{eq:D1bar} (i.e. $\uopt \in \Uopt$) on the original network. We prove this part by mathematical induction: First, if the network has a single edge $\e$, then $\tollopt_e=\lambda^*_e$ is the toll price vector. Second, if the network is parallel, then $\tollopt_e=\lambda^*_e$ for all parallel edges $\e \in \E$ is the toll price vector. Third, if the argument holds on two series parallel networks $G^1$ and $G^2$, then there exist vectors $\tau^{1*}$ and $\tau^{2*}$ such that $\(\uopt, \toll^{1*}\)$ and $\(\uopt, \toll^{2*}\)$ are optimal solutions of \eqref{eq:D1bar} restricted on the sub-network $G^1$ and $G^2$, respectively. Then, we can check that $\tollopt=\(\toll^{*1}, \toll^{2*}\)$ is feasible in \eqref{eq:D1bar} and achieves the same objective function as the sum of that restricted in each one of the two sub-networks when the two networks are connected in series or in parallel. From Lemma \ref{lemma:FF}, we know that the optimal values of the both dual problems equal to the optimal social welfare given $\xopt$. Thus, $\(\uopt, \tollopt\)$ is also an optimal solution of \eqref{eq:D1bar}. 
\QEDA

\vspace{0.3cm}
\noindent\emph{Proof of Lemma \ref{lemma:utility_walrasian}.} For any $\uopt \in \Uopt$, we define $\lambda^* = \(\lambda^*\)_{\r \in \R}$ as follows: 
\begin{align*}
    \lambda_r^* = \max_{\ball \in \Ball} \W_r(\ball) - \sum_{\m \in \ball} \umopt, \quad \forall \r \in \Ropt.  
\end{align*}
Analogous to the proof of Lemma \ref{lemma:equivalent_economy}, we can show that $\(\yopt, \uopt\)$ is a Walrasian equilibrium if and only if $\(\yopt, \uopt, \lambda^*\)$ satisfies the feasibility constraints of \eqref{eq:LP2k} and \eqref{eq:D2k} and the complementary slackness conditions. Therefore, $\(\uopt, \lambda^*\)$ must be an optimal solution of \eqref{eq:D2k} and $\uopt \in \Uopt$.  \QEDA
\end{appendix}
\end{document}